\documentclass[fp]{jpsj3}
\usepackage{txfonts}
\usepackage{multirow}
\usepackage{bm}
\usepackage{amsmath}
\usepackage{pdfpages}
\usepackage{mathrsfs}

\def\pdv#1#2{\frac{\partial #1}{\partial #2}}
\def\ket#1{| #1 \rangle}

\title{Superconducting Diode Effect in Double Quantum Dot Device}

\author{Go Takeuchi and Mikio Eto}
\inst{Faculty of Science and Technology, Keio University, Hiyoshi,
Yokohama 223-8522, Japan} 

\abst{Superconducting diode effect (SDE) is theoretically examined in
double quantum dot coupled to three superconducting leads,
$L$, $R1$ and $R2$. Lead $L$ is commonly connected to two
quantum dots (QD1, QD2)
while lead $R1$ ($R2$) is connected to QD1 (QD2) only.
The phase differences $\varphi_{1}$ between leads $L$ and $R1$ and
$\varphi_{2}$ between leads $L$ and $R2$ are tuned independently.
The critical current into lead $R1$ depends on its direction unless
$\varphi_{2} = 0$, $\pi$, which is ascribable to the formation of
Andreev molecule between the QDs.
In the absence of electron-electron interaction $U$ in the QDs,
the spectrum of the Andreev bound states forms Dirac cones in the
$\varphi_{1}-\varphi_{2}$ plane if the energy levels in the QDs
are tuned to the Fermi level in the leads. The SDE is enhanced 
to almost 30\% when $\varphi_{2}$ is set to the value at
the Dirac points.
In the presence of $U$, the SDE is still observed
when $U$ is smaller than the superconducting energy gap in the leads.
Our device should be one of the minimal models for the SDE
since a similar device with a single QD does not show the SDE.
}

\begin{document}
\maketitle

\section{Introduction}

Superconducting diode effect (SDE) has been studied intensively in various systems.
The critical current can depend on the direction when both the time-reversal and
spatial-inversion symmetries are broken. This effect could be applied to the
dissipationless rectification in superconducting circuits, corresponding to
pn junctions in traditional electronics.
The SDE was observed, e.\ g., in a superconductor without the
spatial-inversion symmetry in a magnetic field.\cite{Ando2020}
Besides the intrinsic SDE in superconductors, the SDE was proposed in
Josephson junction devices, so-called Josephson
diode:\cite{Hu2007,Reynoso2008,Zazunov2009} e.g.,
a semiconductor nanowire connected to two superconducting leads in the
presence of spin-orbit interaction and Zeeman
effect.\cite{Yokoyama2013,Yokoyama2014} The Josephson diodes were realized
using various materials,\cite{Baumgartner2022,Wu2022,Pal2022,Costa2023,
Banerjee2023} asymmetric superconducting quantum interference devices
(SQUIDs),\cite{Ciaccia2023,Greco2023,Gupta2023}
multi-terminal devices,\cite{Gupta2023,Coraiola2023} etc.
 
Recently, Matsuo {\it et al}.\ observed the SDE in coupled Josephson junctions
consisting of three superconducting leads: Each junction is connected to
a common lead (lead $L$ with superconducting phase $\phi_L$) and another
lead ($R1$ or $R2$ with $\phi_{R1}$, $\phi_{R2}$,
respectively).\cite{Matsuo2023_1,Matsuo2023_2}
The former gives rise to the coherent coupling between the junctions by the
crossed Andreev reflection, in which an electron from a junction is reflected
to a hole in another and vice versa, forming an
``Andreev molecule.''\cite{Kornich2019,Kornich2020,Matsuo2022,
Pillet2023,Matsuo2023_3}
Then, the supercurrent to lead $R1$, $I_1$,  depends on not only
$\varphi_{1}=\phi_{L}-\phi_{R1}$ but also
$\varphi_{2}=\phi_{L}-\phi_{R2}$.
Hence $I_1$ can be controlled by phase $\varphi_{2}$ in the nearby
superconductor. For a given $\varphi_2$ ($\ne 0$, $\pi$),
$I_1$ flows at $\varphi_{1}=0$ (anomalous Josephson effect)\cite{Matsuo2023_2}
and shows the SDE.\cite{Matsuo2023_1}
The SDE in this device does not require the spin-orbit interaction.

In this paper, we theoretically examine the SDE in three-terminal Josephson
junctions using semiconductor quantum dots (QDs).
The QDs are useful devices to examine various transport phenomena due to their
tunability of various parameters, particularly discrete energy levels
in the QDs.
The QDs connected to the superconducting leads were studied in various
contexts, including the competition between the Kondo effect and formation of
the Andreev bound state, Yu-Shiba-Rusinov states  (see Ref.\ \citen{Meden2019}
for the review), and observation of $\pi$-junction.\cite{Dam2006}
The double quantum dot (DQD) was also studied,\cite{Choi2000,Pan2006,Wang2011}
for the application to the Cooper pair splitter to create an entangled pair of
electrons,\cite{Recher2001,Deacon2015}
observation of the poor man's Majorana states as a minimal
Kitaev chain,\cite{Leijnse2012,Dvir2023,Bordin2023} etc.

\begin{figure}
\centering
\includegraphics*[width=0.45\textwidth]{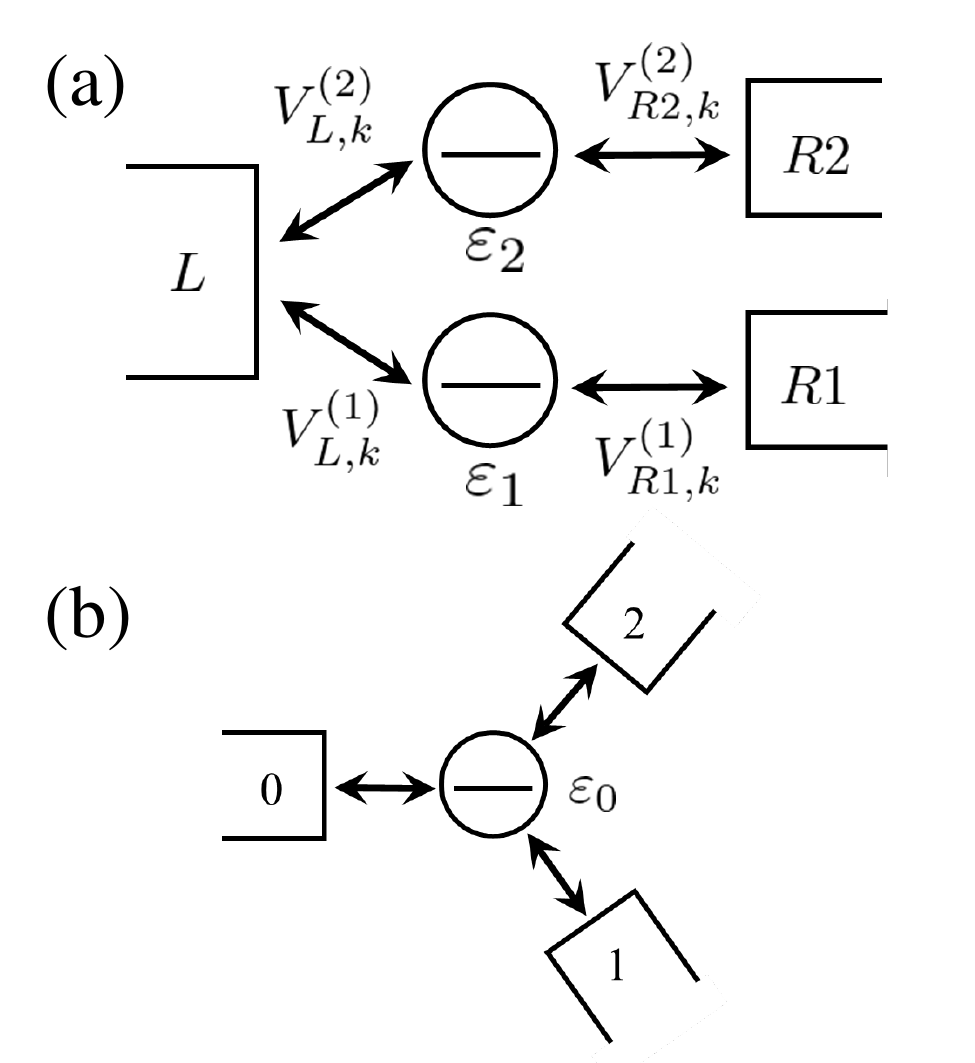}
\caption{(a) Our model of double quantum dot connected to three
superconducting leads. QD1 (QD2) with energy level $\varepsilon_1$
($\varepsilon_2$) is connected to leads $L$ and $R1$ ($R2$).
The strength of the tunnel couplings is characterized by the
linewidth functions in Eqs.\ (\ref{eq:linewidth_fn_R}) and
(\ref{eq:linewidth_fn_L}). The phase differences are given by
$\varphi_1=\phi_L-\phi_{R1}$ and
$\varphi_2=\phi_L-\phi_{R2}$, where
$\phi_L$, $\phi_{R1}$, and $\phi_{R2}$ are the superconducting
phases in respective leads.
(b) Model of a single QD connected to three superconducting
leads. This model is examined in Appendix C.
}
\label{Fig1:Model}
\end{figure}

We focus on the SDE in the DQD connected to three superconducting leads,
depicted in Fig.\ \ref{Fig1:Model}(a),
in a similar geometry to the device by Matsuo
{\it et al}.\cite{Matsuo2023_1,Matsuo2023_2}.
The Andreev molecule is formed between the QDs by the crossed Andreev
reflection at lead $L$, which makes the supercurrent $I_1$ to lead $R1$
depend on both the phases $\varphi_1$ and $\varphi_2$.
We consider single energy levels in the QDs, assuming that the level spacing
is larger than the superconducting gap $\Delta_0$ in the leads. We also assume
that their level broadening and Coulomb interaction $U$ in the QDs are much
smaller than $\Delta_0$, which simplifies the analysis of the supercurrent. 
The aim of this study is to elucidate the condition for the enhanced SDE.

First, we examine the case without $U$ in the QDs. We calculate the
energies of the Andreev bound states as a function of $\varphi_{1}$ and
$\varphi_{2}$ by solving the Bogoliubov-de Gennes (BdG) equation.
When the energy levels in the QDs are tuned to match the Fermi level
in the leads, the
energies become zero at some $(\varphi_{1},\varphi_{2})$ and form Dirac
cones around the points in the $\varphi_{1}-\varphi_{2}$ plane.
When $\varphi_{2}$ is fixed at the value of the Dirac points,
the supercurrent $I_1$ shows a large SDE. Its efficiency can be
around 30\%.
Second, we take $U$ into account. Then, the groundstate is either
spin singlet or doublet, depending on the phases $\varphi_{1}$ and
$\varphi_{2}$. The latter appears in the vicinity of the Dirac
points in the case of $U=0$.
The supercurrent flows even in the doublet phase in our device.
The transition between the spin states enhances the SDE.

Note that (i) the Dirac points were previously examined as Weyl singularities
in the Josephson junctions with four or more superconducting
leads.\cite{Yokoyama2015,Riwar2016,Klees2020}
In our device, the Dirac points can be realized by tuning the QDs.
(ii) We find the anomalous Josephson effect and Dirac cones, but
do not the SDE in a single QD connected to three superconducting leads,
as depicted in Fig.\ \ref{Fig1:Model}(b). Hence, our model in
Fig.\ \ref{Fig1:Model}(a) should be one of the minimal models for the SDE
as long as simple QDs are considered without the chirality,\cite{Cheng2023}
spin-orbit interaction,\cite{Debnath2024} etc.

This paper is organized as follows.
We explain our model and calculation methods in the absence and presence of
$U$ in Sect.\ 2. Section 3 is devoted to the calculated results in the
absence of $U$. We show that the spectrum of the Andreev bound states
forms the Dirac cones in the $\varphi_{1}-\varphi_{2}$ plan.
The supercurrent to lead $R1$ and
its SDE are enhanced when $\varphi_{2}$ is chosen to be the value
at the Dirac points. In Sect.\ 4, we show the
calculated results in the presence of $U$, showing a large SDE
even in this case. We give the conclusions and discussion in Sect.\ 5.
For comparison, we examine a single QD device in
Fig.\ \ref{Fig1:Model}(b) in Appendix C.

\section{Model and Calculation Method}

\subsection{Model and effective Hamiltonian}

Our model of the DQD is depicted in Fig.\ \ref{Fig1:Model}(a).
One of the QDs (QD1) is tunnel-coupled
to superconducting leads $L$ and $R1$ while the other (QD2) is to $L$ and
$R2$. The Hamiltonian is given by
\begin{equation}
    H^{\mathrm{DQD}}=H_{\mathrm{dots}}+H_{\mathrm{S}}+H_{\mathrm{T}}
\end{equation}
with
\begin{align}
  H_{\mathrm{dots}}
&=
  \sum_{j=1, 2}\left(
  \sum_{\sigma}\varepsilon_{j} d_{j, \sigma}^{\dag}d_{j, \sigma}
  +Un_{j \uparrow}n_{j \downarrow}\right),
\\
  H_{\mathrm{S}}
&=
  \sum_{\alpha,k}
  \left[\sum_{\sigma}
  \varepsilon_{k} c_{\alpha, k\sigma}^{\dag}c_{\alpha, k\sigma}
  -\left(
  \Delta_{\alpha}c_{\alpha, k\uparrow}^{\dag}c_{\alpha, -k\downarrow}^{\dag}
  +\mathrm{h.c.}\right) \right],
\\
  H_{\mathrm{T}}
&=
  \sum_{j=1, 2}\sum_{k\sigma}\left[
  \left(V_{L, k}^{(j)}c_{L, k\sigma}^{\dag}+V_{Rj, k}^{(j)}
  c_{Rj, k\sigma}^{\dag}\right)d_{j, \sigma}+\mathrm{h.c.}\right],
\end{align}
where $c_{\alpha, k\sigma}^{\dag}$ and $c_{\alpha, k\sigma}$ are
creation and annihilation operators for
an electron in lead $\alpha$ $(=L, R1, R2)$ with state $k$ and
spin $\sigma$, whereas $d_{j, \sigma}^{\dag}$ and $d_{j, \sigma}$
are those in QD$j$ with spin $\sigma$.
$n_{j,\sigma}=d_{j, \sigma}^{\dag} d_{j, \sigma}$ is the number
operator in QD$j$ with spin $\sigma$.
In $H_{\mathrm{dots}}$, $\varepsilon_{j}$ is the energy level
in QD$j$ ($j=1, 2$) and $U$ is the electron-electron interaction
in the QDs.
The superconducting leads are described by $H_{\mathrm{S}}$,
where $\Delta_{\alpha}=\Delta_0e^{i\phi_{\alpha}}$ with
the s-wave superconducting gap $\Delta_0$ and phase $\phi_{\alpha}$
in lead $\alpha$.

The tunnel couplings are given by $H_{\mathrm{T}}$ in which
$V_{\alpha, k}^{(j)}$ are taken to be positive.
The strength of the tunneling between QD$j$ and lead
$Rj$ is characterized by the linewidth functions,
\begin{equation}
  \Gamma_{Rj}(\varepsilon)=\pi \sum_k
\left[ V_{Rj, k}^{(j)} \right]^2  \delta(\varepsilon-\varepsilon_k)
\label{eq:linewidth_fn_R}
\end{equation}
($j=1, 2$) while that between the DQD and lead $L$ is by
\begin{equation}
  \Gamma_{L;ij}(\varepsilon)=\pi \sum_kV_{L, k}^{(i)}V_{L, k}^{(j)}
  \delta(\varepsilon-\varepsilon_k)
\label{eq:linewidth_fn_L}
\end{equation}
in the form of $2 \times 2$ matrix ($i,j=1, 2$). 
We assume their weak $\varepsilon$-dependence around the
Fermi level $E_{\mathrm F}$ ($=0$)
and simply express $\Gamma_{Rj}$ and $\Gamma_{L;ij}$
for $\varepsilon \simeq E_{\mathrm F}$. The level broadening
in QD$j$ is given by $\Gamma_{Rj}+\Gamma_{L;jj}$.
For the off-diagonal elements in Eq.\
(\ref{eq:linewidth_fn_L}),
we introduce a parameter $p$ ($0\le p\le 1$) by
\begin{equation}
  \Gamma_{L;12}=\Gamma_{L;21}=\sqrt{\Gamma_{L;11}\Gamma_{L;22}}p,
  \label{eq:parameter_p}
\end{equation}
which determines the coherent coupling between the
QDs through lead $L$.\cite{Kubo2006,Zhang2022,Zhang2024}
$p$ is identical to the overlap integral between the conduction
modes coupled to QD1 and QD2 in the lead at the energy
$\varepsilon \simeq E_{\mathrm F}$.\cite{Eto2020}
In the case of single channel in the lead, $p=1$ and the
connection between the QDs is the maximal.
With an increase in the channel number, $p$ usually decreases.
For $p=0$, two QDs become independent of each other.
In experiments, $p$ is determined by the device structure.

In this paper, we restrict ourselves to the low-energy regime compared
with the superconducting gap,
assuming that $|\varepsilon_j|$,
$U$, $\Gamma_{Rj}$, $\Gamma_{L;jj} \ll \Delta_0$ for $j=1,2$. By the
Schrieffer-Wolff transformation,\cite{Bravyi2011,Scherubl2019,Spethmann2024}
we derive the effective Hamiltonian to the second-order of $H_{\mathrm{T}}$,
\begin{equation}
  \label{eq:hamilotonian_DQD}
  H_{\mathrm{eff}}^{\mathrm{DQD}}
 =H_{\mathrm{dots}}+H_{\mathrm{LAR}}+H_{\mathrm{CAR}}.
\end{equation}
Here,
\begin{align}
  H_{\mathrm{LAR}}
&=
  \sum_{j=1, 2} \left(
  \Gamma_{\mathrm{LAR}, j}d_{j \uparrow}^{\dag}d_{j \downarrow}^{\dag}
  +\mathrm{h.c.}\right),
\\
  H_{\mathrm{CAR}}
&=
  \Gamma_{\mathrm{CAR}} \left(
  d_{1 \uparrow}^{\dag}d_{2 \downarrow}^{\dag}
 -d_{1 \downarrow}^{\dag}d_{2 \uparrow}^{\dag}
  \right)+\mathrm{h.c.},
\end{align}
where
\begin{align}
  \Gamma_{\mathrm{LAR}, j}
&=
 -\Gamma_{L;jj}e^{i\phi_{L}}-\Gamma_{Rj}e^{i\phi_{Rj}},
\\
  \Gamma_{\mathrm{CAR}}
&=
-\sqrt{\Gamma_{L;11}\Gamma_{L;22}}pe^{i\phi_L}.
\label{eq:Gamma_CAR}
\end{align}
$H_{\mathrm{LAR}}$ represents the local Andreev reflection in which
an electron in a QD is reflected to a hole in the same QD and
vice versa at leads $R1$, $R2$, or $L$, whereas
$H_{\mathrm{CAR}}$ represents the
crossed Andreev reflection between the QDs at lead $L$.
Note that the strength of the latter is proportional to the parameter
$p$ in Eq.\ (\ref{eq:parameter_p}).

\subsection{Calculation method without $U$}

We begin with the case of ``one-body problem'' in the absence of $U$. 
$H_{\mathrm{eff}}^{\mathrm{DQD}}$ in Eq.\ (\ref{eq:hamilotonian_DQD})
is rewritten in the Nambu form as
\begin{equation}
  H_{\mathrm{eff}}^{\mathrm{DQD}}
  =\begin{pmatrix}
    d_{1\uparrow}^{\dag}, d_{1\downarrow}, d_{2\uparrow}^{\dag}, d_{2\downarrow}
    \end{pmatrix}
   \begin{pmatrix}
    \varepsilon_1 & \Gamma_{\mathrm{LAR}, 1} & 0 & \Gamma_{\mathrm{CAR}} \\
    \Gamma_{\mathrm{LAR},1}^* & -\varepsilon_1 & \Gamma_{\mathrm{CAR}}^* & 0 \\
    0 & \Gamma_{\mathrm{CAR}} & \varepsilon_2 & \Gamma_{\mathrm{LAR},2} \\
    \Gamma_{\mathrm{CAR}}^* & 0 & \Gamma_{\mathrm{LAR},2}^* & -\varepsilon_2
  \end{pmatrix}
   \begin{pmatrix}
    d_{1\uparrow} \\
    d_{1\downarrow}^{\dag} \\
    d_{2\uparrow} \\
    d_{2\downarrow}^{\dag}
   \end{pmatrix} +\sum_{j=1, 2}\varepsilon_j.
\label{eq:Hamiltonian_Nambu}
\end{equation}
As shown in Appendix A, the BdG equation yields four eigenenergies,
$\pm E_1$, $\pm E_2$ ($0 \le E_1 \le E_2$), corresponding to the
Andreev bound states. After the Bogoliubov transformation, we obtain
the groundstate energy,
\begin{equation}
  E_{\mathrm{GS}}=-\sum_{n=1,2} E_n +\sum_{j=1, 2}\varepsilon_j.
\label{eq:GSenergy0}
\end{equation}
Note that $E_n$ is a function of the phase differences $\varphi_1$ and
$\varphi_2$, where
\begin{equation}
 \varphi_j=\phi_L-\phi_{Rj}
\label{eq:phase12}
\end{equation}
for $j=1, 2$. We denote the supercurrent into lead $Rj$ by 
$I_j$ ($j=1, 2$), which is given by
\begin{align}
  I_j &= \frac{2e}{\hbar} \pdv{E_{\mathrm{GS}}}{\varphi_j}
\label{eq:Josephson_current0}
\\
   &= -\frac{2e}{\hbar} \sum_{n=1,2}
       \pdv{E_n(\varphi_1,\varphi_2)}{\varphi_j}
\label{eq:Josephson_current}
\end{align}
at temperature $T=0$.
This formula is derived using the Hellman-Feynman theorem.\cite{Meden2019}

\subsection{Calculation method with $U$}

In the presence of $U$, we diagonalize the effective Hamiltonian
$H_{\mathrm{eff}}^{\mathrm{DQD}}$ in Eq.\ (\ref{eq:hamilotonian_DQD})
in the space of many-body states in the DQD. The space is divided into three
subspaces with the total spin $S=0$, $1/2$, and $1$.

In the subspace of spin singlet ($S=0$), there are five states,
$\ket{0}$, $\ket{s}$, $\ket{1\uparrow, 1\downarrow}$,
$\ket{2\uparrow, 2\downarrow}$, and
$\ket{1\uparrow, 1\downarrow, 2\uparrow, 2\downarrow}$, where
$\ket{0}$ is the empty state in the DQD,
$\ket{j, \sigma}=d_{j, \sigma}^{\dag}\ket{0}$, and
\begin{equation}
  \ket{s}=  \frac{1}{\sqrt{2}}
  \left(\ket{1\uparrow, 2\downarrow}-\ket{1\downarrow, 2\uparrow}\right).
\label{eq:singlet_s}
\end{equation}
Using these states as a basis set, the Hamiltonian is represented by
Eq.\ (\ref{eq:matrix_singlet}) in Appendix B.
Its diagonalization yields the energies of the many-body Andreev bound states.
The lowest one is denoted by $E_{\mathrm{singlet}}$.

The subspace of spin doublet ($S=1/2$) consists of the space with $S_z=1/2$
and that with $S_z=-1/2$. Each space includes four states;
$\ket{1\uparrow}$,
$\ket{2\uparrow}$, $\ket{1\uparrow, 2\uparrow, 2\downarrow}$,
$\ket{1\uparrow, 1\downarrow, 2\uparrow}$ in the former and
$\ket{1\downarrow}$, $\ket{2\downarrow}$,
$\ket{1\downarrow, 2\uparrow, 2\downarrow}$,
$\ket{1\uparrow, 1\downarrow, 2\downarrow}$ in the latter.
The diagonalization of the representation matrix in
Eq.\ (\ref{eq:matrix_doublet}) gives the lowest energy,
$E_{\mathrm{doublet}}$,
which is doubly degenerate with $S_z=\pm 1/2$.

Finally, three spin-triplet states, 
$\ket{1\uparrow, 2\uparrow}$,
\begin{equation}
  \ket{t}=\frac{1}{\sqrt{2}}\left(\ket{1\uparrow, 2\downarrow}
          +\ket{1\downarrow, 2\uparrow}\right),
\label{eq:triplet_t}
\end{equation}
and $\ket{1\downarrow, 2\downarrow}$, are already the eigenstates of
$H_{\mathrm{eff}}^{\mathrm{DQD}}$. The eigenenergies are
$\varepsilon_1+\varepsilon_2 \equiv E_{\mathrm{triplet}}$.

The groundstate energy $E_{\mathrm{GS}}$ is determined by the minimum
among $E_{\mathrm{singlet}}$, $E_{\mathrm{doublet}}$, and
$E_{\mathrm{triplet}}$. Note that both $E_{\mathrm{singlet}}$
and $E_{\mathrm{doublet}}$ depend on the phase differences
$\varphi_1$ and $\varphi_2$ in Eq.\ (\ref{eq:phase12}) while
$E_{\mathrm{triplet}}$ does not.

The supercurrent is given by the formula in Eq.\ (\ref{eq:Josephson_current0}).
It flows in the spin-singlet or doublet phases and does not in the triplet
phase.

\begin{figure}
\centering
\includegraphics*[width=0.85\textwidth]{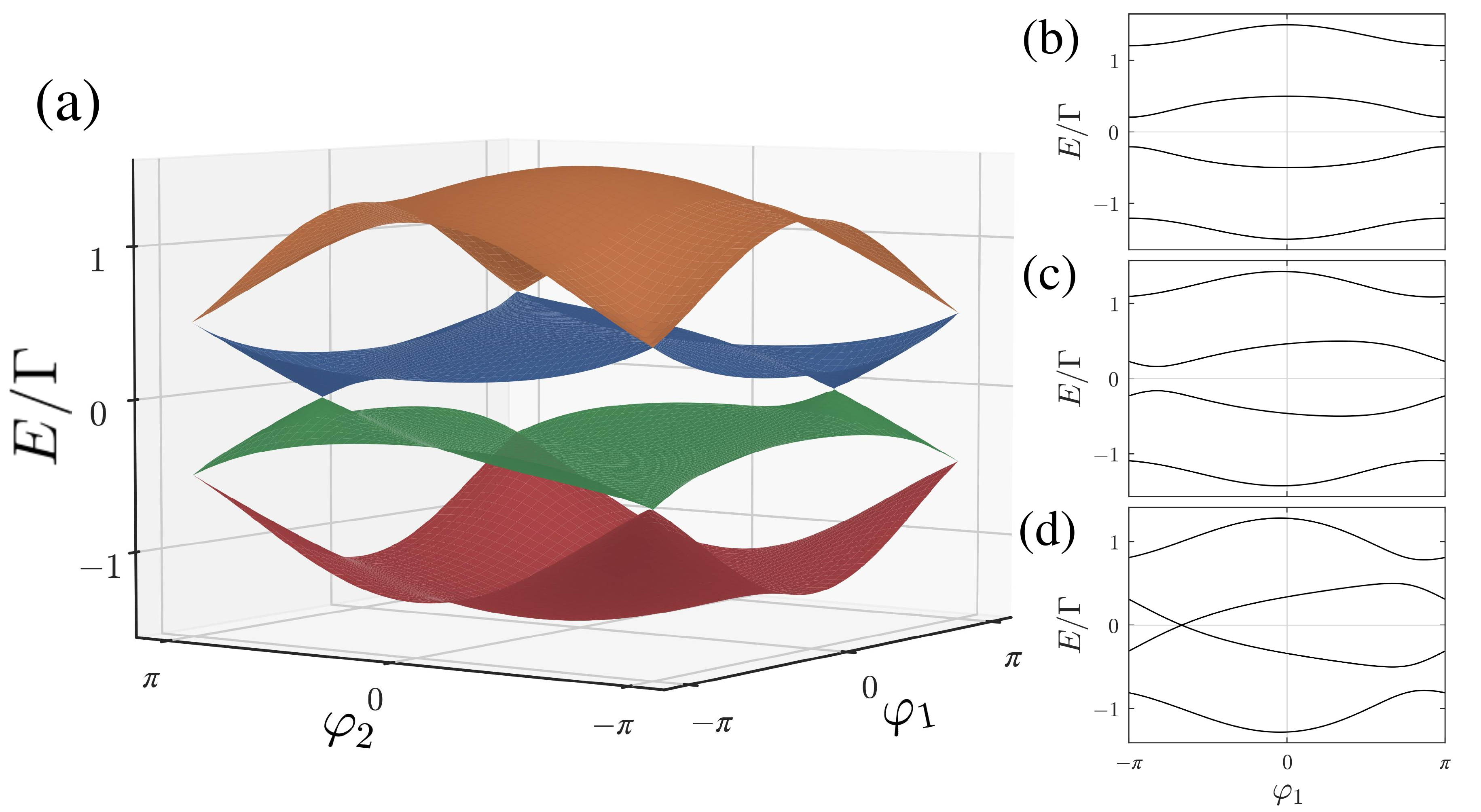}
\caption{(Color online)
Energies of Andreev bound states,
$\pm E_1$ and $\pm E_2$ ($0 \le E_1 \le E_2$),
in the model in Fig.\ \ref{Fig1:Model}(a) with $U=0$.
$\varepsilon_1=\varepsilon_2=0$,
$\Gamma_{L;11}=\Gamma_{R1}=\Gamma_{L;22}=\Gamma_{R2} = \Gamma/2$,
and $p=1$, where $\Gamma$ is the level broadening in the QDs.
(a) The energies plotted in the $\varphi_1-\varphi_2$ plane.
$E_1$ and $-E_1$ form the Dirac cones around the zero points
at $(\varphi_1, \varphi_2)=\pm (2\pi/3, -2\pi/3)$.
(b)--(d) Cross sections of panel (a) at (b) $\varphi_2=0$,
(c) $\pi/3$, and (d) $2\pi/3$.
}
\label{Fig2:AndreevBSs}
\end{figure}

\section{Calculated Results in the Absence of $U$}

In this section, we discuss the case of $U=0$.
First, we examine the spectrum of the Andreev bound states.
Figure \ref{Fig2:AndreevBSs}(a) depicts $\pm E_1$ and $\pm E_2$
($0 \le E_1 \le E_2$) in the $\varphi_1-\varphi_2$ plane.
We match the energy levels in the QDs to the Fermi level,
$\varepsilon_1=\varepsilon_2=0$, and set
$\Gamma_{L;11}=\Gamma_{R1}=\Gamma_{L;22}=\Gamma_{R2}
=\Gamma/2$ and $p=1$, where $\Gamma$ is the level broadening
in the QDs.
In this case, $E_1=0$ at $(\varphi_1, \varphi_2)=\pm (2\pi/3, -2\pi/3)$,
which is  derived in Appendix A. The energies of
$\pm E_1(\varphi_1, \varphi_2)$ form a ``Dirac cone'' around these
zero-points.

Note that $E_{n}(-\varphi_1, \varphi_2) \ne
E_{n}(\varphi_1, \varphi_2)$ unless $\varphi_2=0$ or $\pi$
although the identity of
$E_{n}(-\varphi_1, -\varphi_2)=E_{n}(\varphi_1, \varphi_2)$
always holds by the time-reversal relation.
Figures \ref{Fig2:AndreevBSs}(b)--(d) depict the cross sections
of Fig.\ \ref{Fig2:AndreevBSs}(a) at (b) $\varphi_2=0$,
(c) $\pi/3$, and (d) $2\pi/3$.
The energies $\pm E_1$ and $\pm E_2$ are even functions of
$\varphi_1$ at $\varphi_2=0$, whereas they are asymmetric
about $\varphi_1=0$ at $\varphi_2 \ne 0$.
$E_1$ has a zero point at $\varphi_2 = 2\pi/3$, corresponding
to the Dirac points.

\begin{figure}
\centering
\includegraphics*[width=0.5\textwidth]{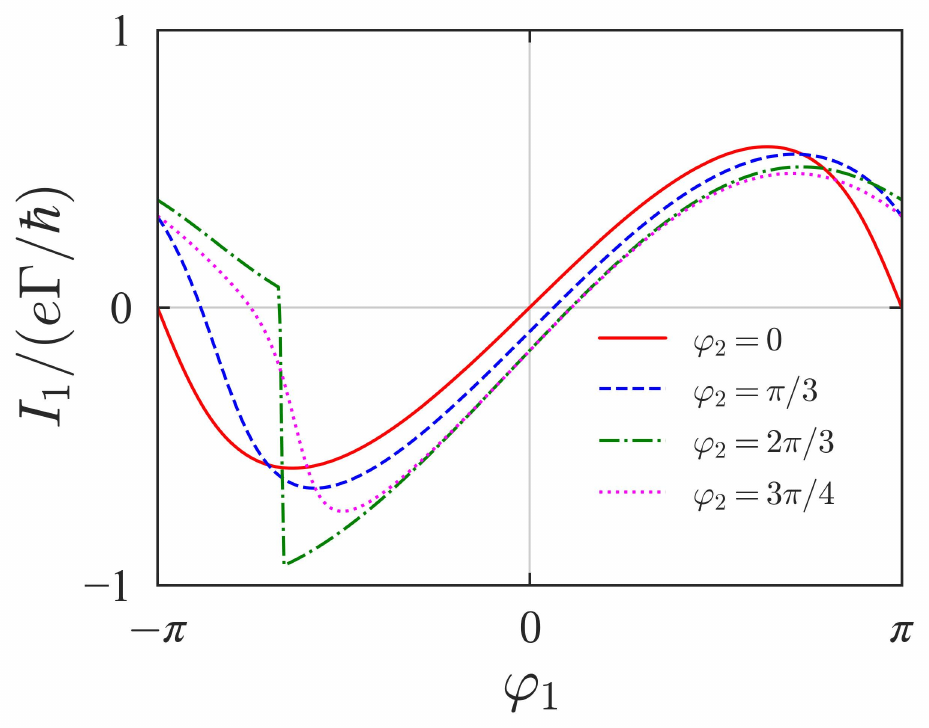}
\caption{(Color online)
Supercurrent $I_1$ to lead $R1$ as a function of
$\varphi_1$, in the model in Fig.\ \ref{Fig1:Model}(a) with $U=0$,
when $\varphi_2$ is fixed at $0$ (solid line),
$\pi/3$ (broken line), $2\pi/3$ (dash-dotted line),
and $3\pi/4$ (dotted line). 
$\varepsilon_1=\varepsilon_2=0$,
$\Gamma_{L;11}=\Gamma_{R1}=\Gamma_{L;22}=\Gamma_{R2}
\equiv \Gamma/2$, and $p=1$.
}
\label{Fig3:current1}
\end{figure}

We calculate the supercurrent into lead $R1$, $I_1$, using the
formula in  Eq.\ (\ref{eq:Josephson_current}).
Figure \ref{Fig3:current1} shows the current as a function of
$\varphi_1$ when $\varphi_2$ is fixed at $0$, $\pi/3$,
$2\pi/3$, and $3\pi/4$.
When $\varphi_2=0$, $I_1$ is positive (negative) at $0< \varphi_1 <\pi$
($-\pi< \varphi_1 <0$). $I_1$ is an odd function of $\varphi_1$, and
hence $I_1=0$ at $\varphi_1 =0$.
When $\varphi_2 \ne 0$ ($0 < \varphi_2 < \pi$), however,
$I_1$ is not an odd function of $\varphi_1$.
$I_1$ flows at $\varphi_1 =0$, indicating an anomalous Josephson effect.
The behavior of $I_1$ changes gradually with $\varphi_2$ in the
positive region of $\varphi_1$, whereas it changes drastically with
$\varphi_2$ in the negative region of $\varphi_1$.
Particularly, $I_1$ shows a step at $\varphi_1 = -2\pi/3$ when
$\varphi_2=2\pi/3$. This is due to the Dirac point since
$\partial E_1/ \partial \varphi_1 <0$ $(>0)$ at
$\varphi_1 = -2\pi/3-0$ $(+0)$.

In Fig. \ref{Fig3:current1}, we observe that the maximum
of $|I_1|$ is different in the positive and negative directions when
$\varphi_2 \ne 0$. If the critical currents
are denoted by $I_{\mathrm{c}}^{\pm}(\varphi_2)=
\max_{\varphi_1}[\pm I_1(\varphi_1, \varphi_2)]$,
$I_{\mathrm{c}}^{-} = I_{\mathrm{c}}^{+}$ for $\varphi_2=0$
and  $I_{\mathrm{c}}^{-} > I_{\mathrm{c}}^{+}$ for
$0 < \varphi_2 < \pi$.
This diode effect is
the maximal when $\varphi_2=2\pi/3$, the value of the
Dirac point. The mechanism is as follows.
$E_1$ is very steep at the Dirac point and thus
$|\partial E_1/ \partial \varphi_1|$ becomes large there
[$\partial E_1/ \partial \varphi_1=\mp \Gamma/4$,
see Appendix A].
Since $\partial E_2/ \partial \varphi_1 >0$ around
the point, the step of $I_1$ is shifted in the negative direction
by Eq.\ (\ref{eq:Josephson_current}), which results in a
large $I_{\mathrm{c}}^{-}$. On the other hand, $I_1$ takes the
maximum in the positive region of $\varphi_1$.
$I_{\mathrm{c}}^{+}$ changes slowly with $\varphi_2$,
as seen above.
In consequence, the SDE is the maximal at $\varphi_2=2\pi/3$.
It should be mentioned that the Dirac cone does not
give rise to the SDE by itself although it is anisotropic, as
shown in Fig.\ \ref{Fig9:DiracCone}.
This is because $\partial E_1/ \partial \varphi_1$ is the same in
magnitude at $\varphi_1 = -2\pi/3 \pm 0$, as discussed in Appendix A.

\begin{figure}
\centering
\includegraphics*[width=0.4\textwidth]{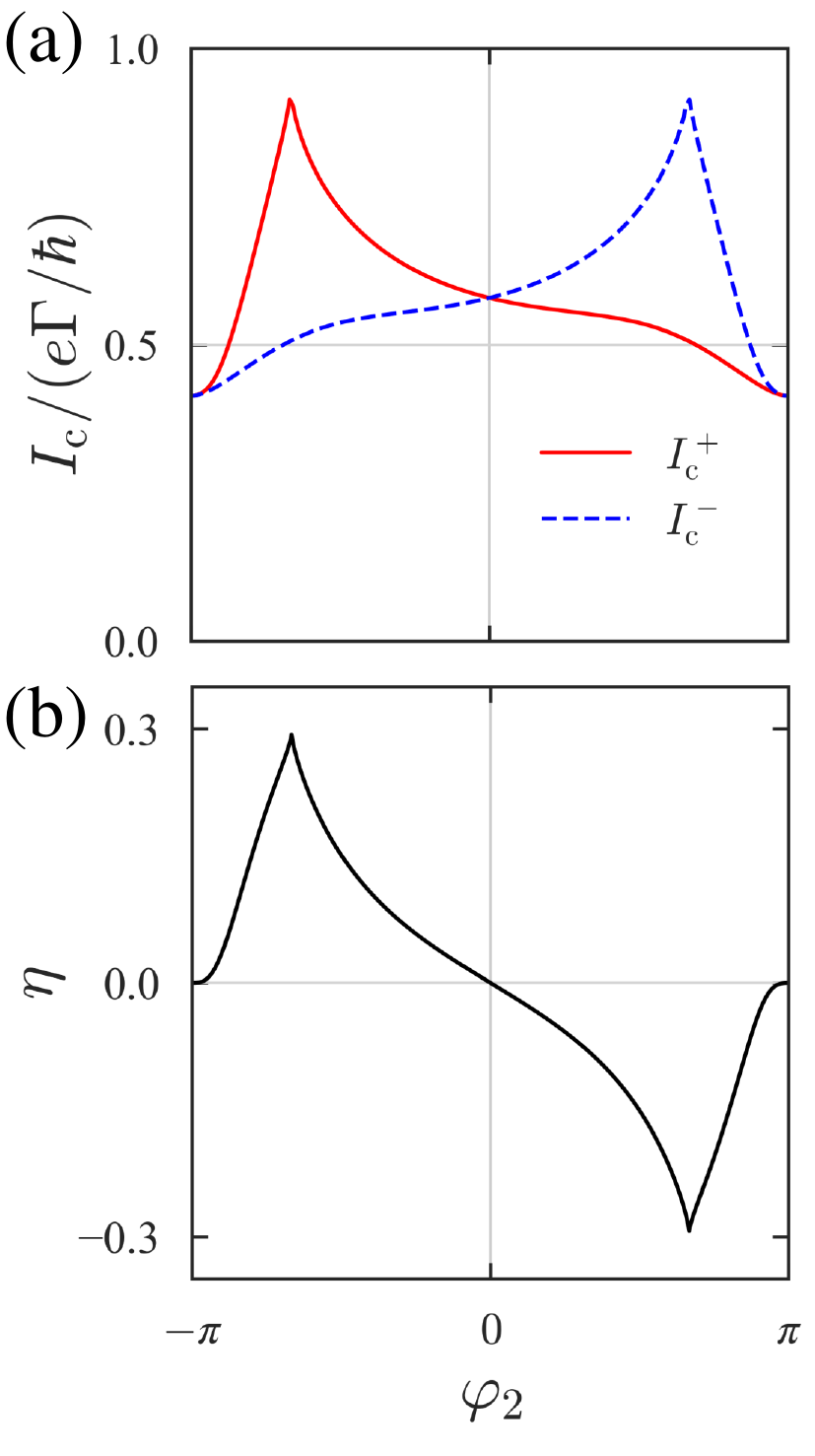}
\caption{(Color online)
(a) Critical current
$I_{\mathrm{c}}^{\pm}$ to lead $R1$ and (b) its
diode efficiency $\eta$ as functions of $\varphi_2$,
in the model in Fig.\ \ref{Fig1:Model}(a) with $U=0$. 
The critical current $I_{\mathrm{c}}^{+}$ ($I_{\mathrm{c}}^{-}$)
in the positive (negative) direction is depicted by solid
(broken) line in panel (a). $\varepsilon_1=\varepsilon_2=0$,
$\Gamma_{L;11}=\Gamma_{R1}=\Gamma_{L;22}=\Gamma_{R2} \equiv \Gamma/2$,
and $p=1$.
}
\label{Fig4:Ic_eta}
\end{figure}

For the quantitative discussion on the SDE, we define its efficiency by
\begin{equation}
  \eta(\varphi_2)=\frac{I_{\mathrm{c}}^+-I_{\mathrm{c}}^-}
  {I_{\mathrm{c}}^++I_{\mathrm{c}}^-}.
\end{equation}
The time-reversal relation leads to the identities,
$I_{\mathrm{c}}^{\pm}(-\varphi_2)=I_{\mathrm{c}}^{\mp}(\varphi_2)$
and thus $\eta(-\varphi_2)=-\eta(\varphi_2)$.
In Fig.\ \ref{Fig4:Ic_eta}, we plot (a) the critical current into lead $R1$,
$I_{\mathrm{c}}^{\pm}(\varphi_2)$, and (b) the diode efficiency
$\eta(\varphi_2)$. Both $I_{\mathrm{c}}^{\pm}$ and $\pm \eta$
are enhanced to the maximum at $\varphi_2=\mp 2\pi/3$,
the values at the Dirac points.
The efficiency can reach almost $\pm 0.3$ there. This is the main result
in this paper.

\begin{figure}
\centering
\includegraphics*[width=0.6\textwidth]{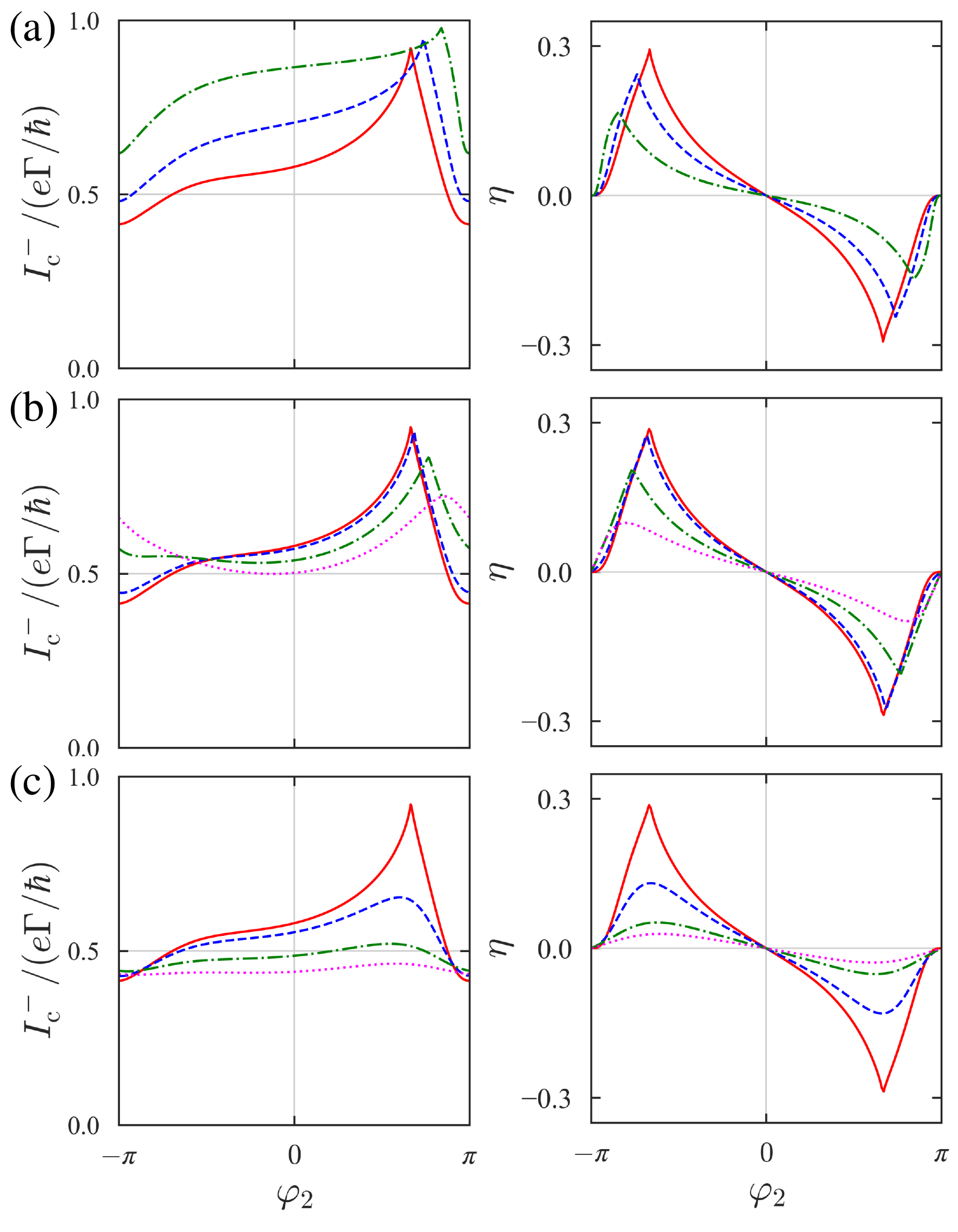}
\caption{(Color online)
Critical current
$I_{\mathrm{c}}^{-}$ to lead $R1$ (left panels) and its
diode efficiency $\eta$ (right panels)
as functions of $\varphi_2$, in the model in Fig.\
\ref{Fig1:Model}(a) with $U=0$.
$\Gamma_{L;11}=\Gamma_{R1}=\Gamma_{L;22}=\Gamma_{R2} \equiv
\Gamma/2$.
(a) $\varepsilon_1=\varepsilon_2=0$ and $p=1$ (solid line),
$0.8$ (broken line), and $0.5$ (dash-dotted line).
(b) [(c)] $\varepsilon_1=-\varepsilon_2 \equiv \varepsilon_0$
[$\varepsilon_1=\varepsilon_2 \equiv \varepsilon_0$] with
$p=1$. $\varepsilon_0/\Gamma=0$ (solid line), $0.15$ (broken line),
$0.35$ (dash-dotted line), and $0.5$ (dotted line).
}
\label{Fig5:Ic_eta2}
\end{figure}

We discuss the generality of the enhancement of the SDE.
In Fig.\ \ref{Fig5:Ic_eta2}, we plot
$I_{\mathrm{c}}^{-}(\varphi_2)$ and $\eta(\varphi_2)$
when (a) $\varepsilon_1=\varepsilon_2=0$ with various $p$,
(b) $\varepsilon_1=-\varepsilon_2 \equiv \varepsilon_0$ with
$p=1$, and (c) $\varepsilon_1=\varepsilon_2 \equiv \varepsilon_0$ with
$p=1$. We choose
$\Gamma_{L;11}=\Gamma_{R1}=\Gamma_{L;22}=\Gamma_{R2} = \Gamma/2$.
[We do not plot $I_{\mathrm{c}}^{+}(\varphi_2)$ since it is given
by $I_{\mathrm{c}}^{-}(-\varphi_2)$.]

With decreasing $p$, the crossed Andreev reflection becomes less
effective [see $\Gamma_{\mathrm{CAR}}$ in Eq.\ (\ref{eq:Gamma_CAR})]
and thus the diode efficiency decreases, as shown in
Fig.\ \ref{Fig5:Ic_eta2}(a). With $p=0.5$,
we still observe the enhanced critical current and
$\eta \simeq 0.15$ around the Dirac point. Note that the position
of the Dirac points moves with $p$, following Eqs.\ (\ref{eq:dirac1})
and (\ref{eq:dirac2}) in Appendix A,
where $I_{\mathrm{c}}^{-}$ ($\eta$) has a cusp (cusps).

When $\varepsilon_1=-\varepsilon_2 \equiv \varepsilon_0$,
the Dirac points exist when $|\varepsilon_0|<\Gamma p/2$ (Appendix A).
As seen in Fig.\ \ref{Fig5:Ic_eta2}(b), 
$I_{\mathrm{c}}^{-}$ (or $I_{\mathrm{c}}^{+}$) and
$|\eta|$ are the largest at the points
unless $|\varepsilon_0|/\Gamma$ is too large.\cite{com1}
For $|\varepsilon_0|/\Gamma=1/2$, we find an enhancement of
$I_{\mathrm{c}}^{-}$ and $|\eta|$ without a cusp although the
Dirac point just disappears.

When $\varepsilon_1=\varepsilon_2 \equiv \varepsilon_0 \ne 0$,
the Dirac points are absent. We still observe the SDE
with $|\eta| \simeq 0.13$ for $\varepsilon_0/\Gamma=0.15$
in Fig.\ \ref{Fig5:Ic_eta2}(c).
The SDE becomes less prominent for $\varepsilon_0/\Gamma \gtrsim 0.35$.

We comment on the existence of Dirac points with
arbitrary values of $\Gamma_{L;11}$, $\Gamma_{R1}$,
$\Gamma_{L;22}$, and $\Gamma_{R2}$.
When $\varepsilon_1=\varepsilon_2 =0$,
the Dirac points exist if $b=0$ in Eq.\ (\ref{eq:equation_of_b})
at some $(\varphi_1,\varphi_2)$. This condition is usually satisfied
when $\Gamma$'s are in the same order of magnitude.
In the case of $p=1$,
the condition is given by a triangle inequality in Eq.\
(\ref{eq:triangle_inequality}).

In summary, the SDE of the supercurrent to lead $R1$
is largely enhanced when $\varphi_2$ is set to the value at the
Dirac points when they are present. A large SDE is still observable
when the Dirac points are absent and thus the Dirac cone becomes
the spectrum of a ``massive particle.''
Around the dips of $E_1(\varphi_1,\varphi_2)$,
$\partial E_1/ \partial \varphi_1$ takes large values in magnitude
with both positive and negative signs. This enhances the critical
current $I_{\mathrm{c}}^{-}$ ($I_{\mathrm{c}}^{+}$) if
$\partial E_2/ \partial \varphi_1 >0$ ($<0$) around the dips of
$E_1$, which results in a large SDE.

\section{Calculated Results in the Presence of $U$}

In the presence of $U$, we choose the energy levels in the QDs
to satisfy the electron-hole symmetry, $\varepsilon_1=\varepsilon_2=-U/2$.
We set $\Gamma_{L;11}=\Gamma_{R1}=\Gamma_{L;22}=\Gamma_{R2}
= \Gamma/2$ and $p=1$, with $\Gamma$ being the level broadening
in the QDs.

\begin{figure}
\centering
\includegraphics*[width=0.5\textwidth]{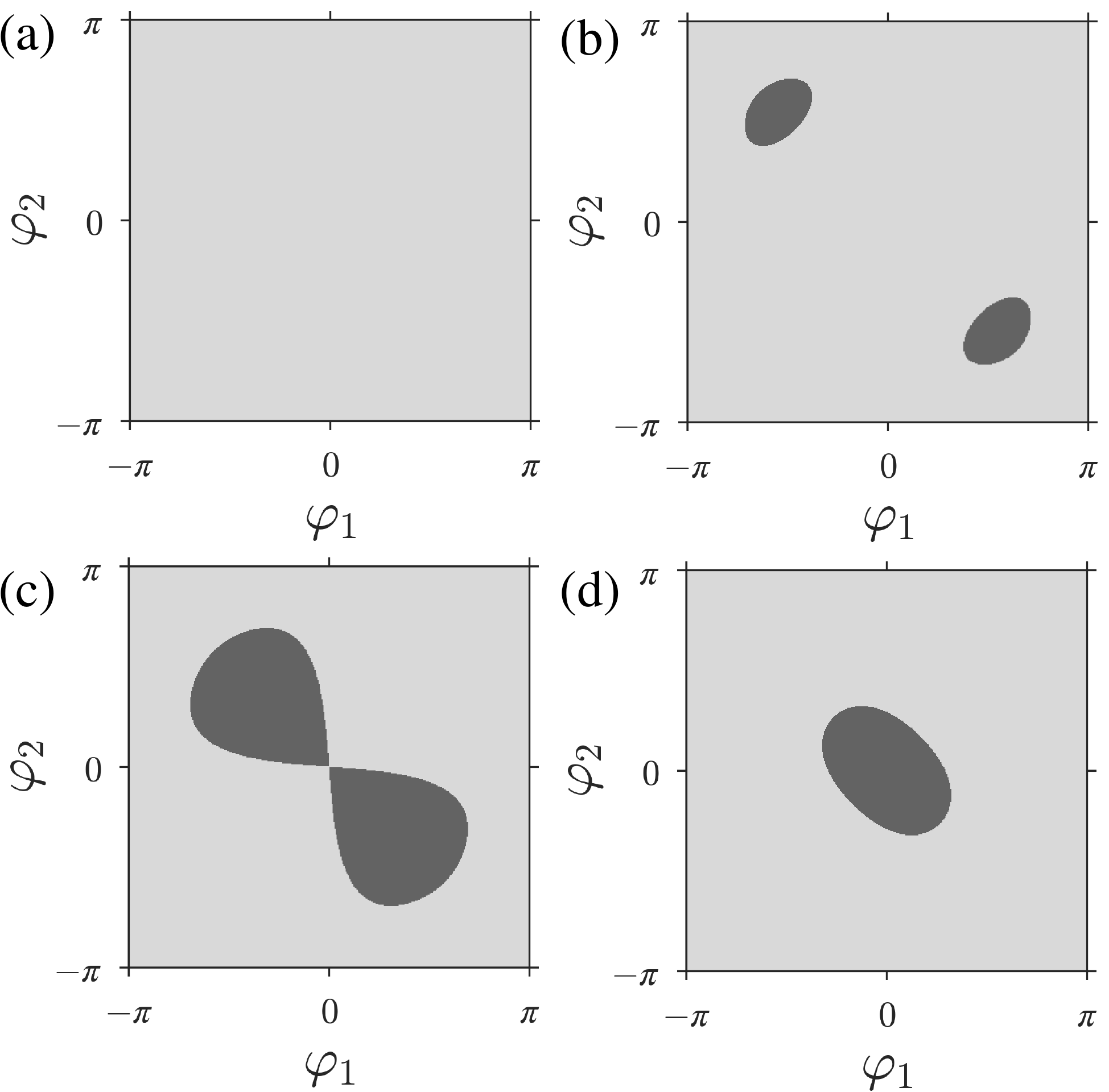}
\caption{
Spin of the groundstate in the model in Fig.\ \ref{Fig1:Model}(a).
The spin-singlet (doublet) regions are indicated by
light gray (dark gray) in the $\varphi_1-\varphi_2$ plane.
$\varepsilon_1=\varepsilon_2=-U/2$,
$\Gamma_{L;11}=\Gamma_{R1}=\Gamma_{L;22}=\Gamma_{R2} \equiv \Gamma/2$,
and $p=1$. (a) $U/\Gamma=0$, (b) $0.5$, (c) $1$, and (d) $2$.
}
\label{Fig6:SingletDoubletRegions}
\end{figure}

The groundstate is either spin singlet or doublet in the presence of $U$.
In Fig.\ \ref{Fig6:SingletDoubletRegions},
we plot the phase diagram in the $\varphi_1-\varphi_2$ plane for
(a) $U/\Gamma=0$, (b) $0.5$, (c) $1$, and (d) $2$.
When $U=0$, the spin-singlet phase is always realized (singlet and
doublet energies are degenerate just at the Dirac points).
For small $U/\Gamma$, the doublet phase appears around the Dirac
points in the case of $U=0$. The regions of the doublet phase are extended
with an increase in $U$ and merge at $U/\Gamma \ge 1$. With increasing
$U$ beyond $U/\Gamma = 2$, the region is shrunk and finally disappears.
(When $U/\Gamma \gg 1$,
$E_{\mathrm{singlet}}\lesssim E_{\mathrm{triplet}} < E_{\mathrm{doublet}}$.
The groundstate is almost $\ket{s}$ in Eq.\ (\ref{eq:singlet_s}) in which each
QD accommodates one electron. Small amplitudes of the other components,
such as $\ket{0}$ and $\ket{1\uparrow, 1\downarrow}$,
weaken the Andreev reflection and reduce the supercurrent.)

\begin{figure}
\centering
\includegraphics*[width=0.7\textwidth]{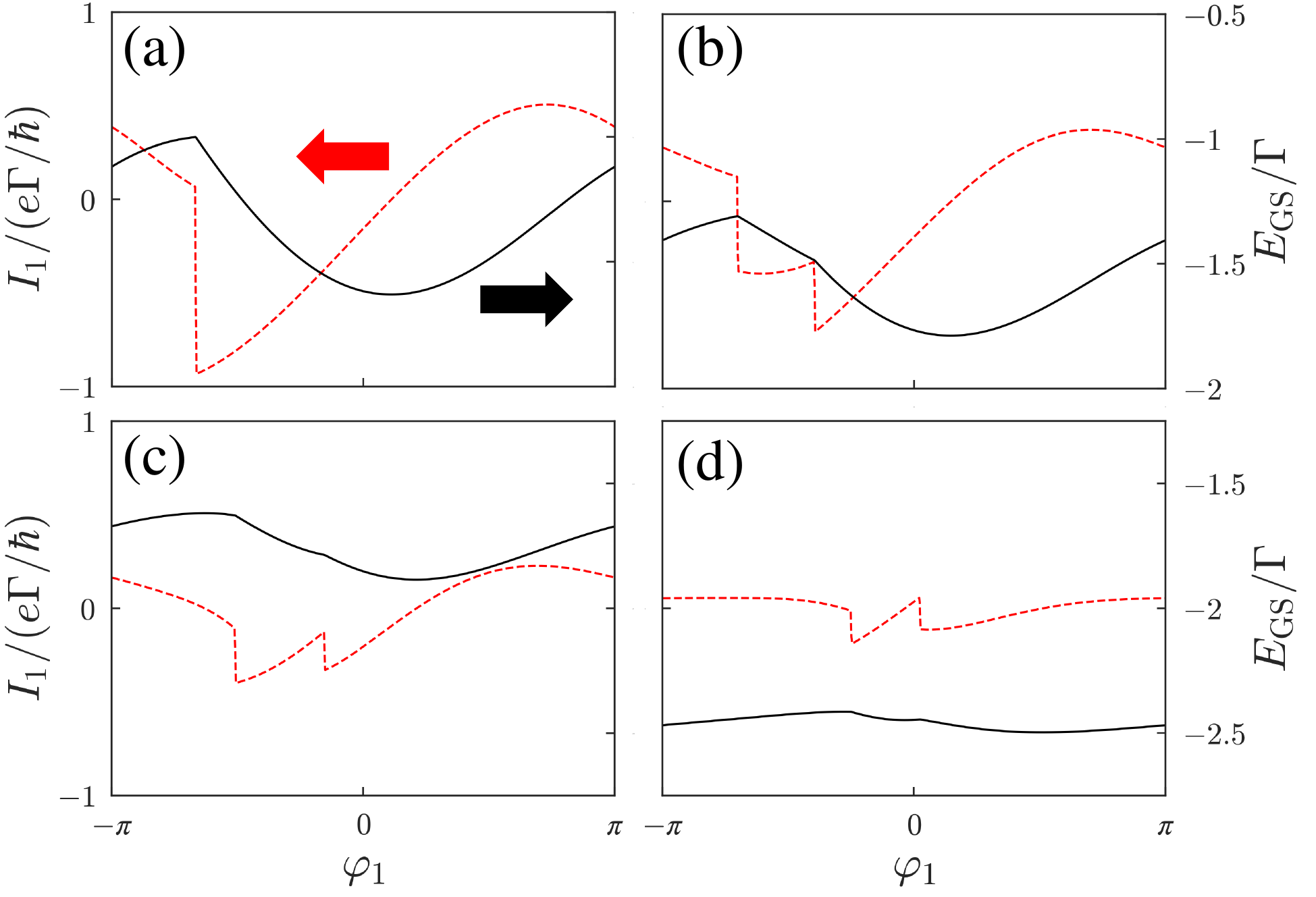}
\caption{(Color online)
Supercurrent $I_1$ to lead $R1$ 
(broken line) and groundstate energy $E_{\mathrm{GS}}$ (solid line),
as functions of $\varphi_1$ with a given $\varphi_2$,
in the model in Fig.\ \ref{Fig1:Model}(a).
$\varepsilon_1=\varepsilon_2=-U/2$,
$\Gamma_{L;11}=\Gamma_{R1}=\Gamma_{L;22}=\Gamma_{R2} \equiv \Gamma/2$,
and $p=1$.
(a) $U/\Gamma=0$, (b) $0.5$, (c) $1$, and (d) $2$.
$\varphi_2$ is chosen at the maximum of $-\eta(\varphi_2)$
in Fig.\ \ref{Fig8:eta_U}:
(a) $\varphi_2=2\pi/3$, (b) $0.55\pi$, (c) $0.63\pi$, and
(d) $0.28\pi$. At the kinks of $E_{\mathrm{GS}}$,
the spin state changes in the groundstate, where $I_1$ shows a step.
}
\label{Fig7:current1Egs}
\end{figure}

In Fig.\ \ref{Fig7:current1Egs}, we plot the energy
of the many-body groundstate, $E_{\mathrm{GS}}$,
and supercurrent to lead $R1$, $I_1$ ,
as functions of $\varphi_1$ when $\varphi_2$ is fixed at the maximum of
$-\eta$, as discussed below. The values of $U/\Gamma$ are the same as
in Fig.\ \ref{Fig6:SingletDoubletRegions}.
When $U=0$, $I_1$ shows a step at $\varphi_1=-2\pi/3$, where
$E_{\mathrm{GS}}$ has a kink, corresponding to the Dirac point,
discussed in the previous section. When $U \ne 0$, $I_1$ shows two steps
at the kinks of $E_{\mathrm{GS}}$, where the groundstate changes
between spin singlet and doublet. Note that the supercurrent flows
even in the doublet phase in our device, as noted in Sect.\ 2.3.
At one of the steps, $-I_1$ takes the maximum, which results in a
large critical current $I_{\mathrm c}^-$ in the negative direction.

\begin{figure}
\centering
\includegraphics*[width=0.6\textwidth]{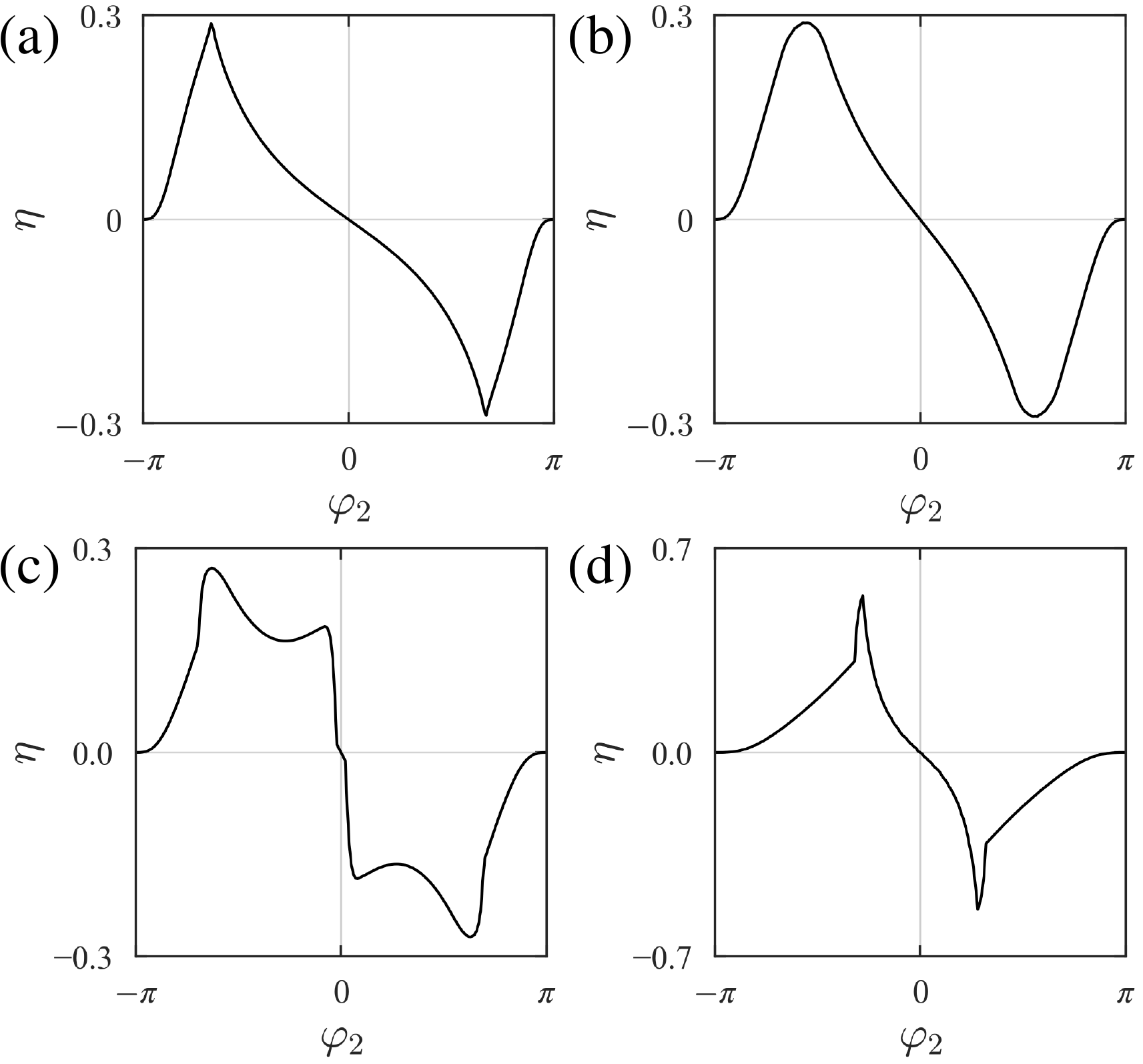}
\caption{
Diode efficiency $\eta$ of the supercurrent to lead $R1$
as a function of $\varphi_2$,
in the model in Fig.\ \ref{Fig1:Model}(a).
$\varepsilon_1=\varepsilon_2=-U/2$,
$\Gamma_{L;11}=\Gamma_{R1}=\Gamma_{L;22}=\Gamma_{R2} \equiv \Gamma/2$,
and $p=1$. (a) $U/\Gamma=0$, (b) $0.5$, (c) $1$, and (d) $2$.
}
\label{Fig8:eta_U}
\end{figure}

Finally, Fig.\ \ref{Fig8:eta_U} indicates the diode efficiency of the
supercurrent $I_1$ as a function of
$\varphi_2$. The enhanced SDE is still seen in the presence of $U$.
The maximum of $|\eta|$ is almost the same for $U/\Gamma=0 \sim 1$.
Although $|\eta|$ is larger than 50\% when $U/\Gamma=2$,
the absolute value of the supercurrent $I_1$ is not large in
Fig.\ \ref{Fig7:current1Egs}(d).
For $U/\Gamma>2$, both the critical current and the SDE decrease with
increasing $U$.

In this section, we have shown the calculated results only when
$\Gamma_{L;11}=\Gamma_{R1}=\Gamma_{L;22}=\Gamma_{R2}$
and $p=1$. When $\Gamma_{L;11}$, $\Gamma_{R1}$, $\Gamma_{L;22}$,
and $\Gamma_{R2}$ are not identical but are in the same order
of magnitude,
we find almost the same enhancement of the SDE as in
Fig.\ \ref{Fig8:eta_U} if $p=1$.
With a decrease in the parameter $p$, the efficiency $|\eta|$ decreases.
For $p=0.5$, we still observe $|\eta| \simeq 20$\%.

\section{Conclusions and Discussion}

We have theoretically examined the SDE in the DQD connected to three
superconducting leads, depicted in Fig.\ \ref{Fig1:Model}(a).
The geometry of this device is similar to that by Matsuo
{\it et al}.\cite{Matsuo2023_1,Matsuo2023_2}
The crossed Andreev reflection at lead $L$ results in the formation of
the Andreev molecule between the QDs, which makes
the supercurrent to lead $R1$, $I_1$, depend on both the phase differences
$\varphi_{1}=\varphi_L-\varphi_{R1}$ and
$\varphi_{2}=\varphi_L-\varphi_{R2}$.
We have formulated the energies of the Andreev bound states
and evaluated the supercurrent $I_1$ in both the cases of $U=0$
and $U \ne 0$ in the QDs.

In the absence of $U$, we have obtained the Andreev bound
states ($\pm E_1$, $\pm E_2$ with $0 \le E_1 \le E_2$) 
by solving the BdG equation.
When the energy levels in the QDs are tuned to match the Fermi level
in the leads, the energies $\pm E_1$ form Dirac cones in the
$\varphi_{1}-\varphi_{2}$ plane.
When $\varphi_{2}$ is fixed at the value of the Dirac points,
one of the critical currents $I_{\mathrm{c}}^{\pm}$ in the
positive or negative directions is enhanced, which leads to a
large SDE. Its efficiency can be around 30\%.
In the absence of Dirac points, an enhanced SDE is
still observable around the dips of $E_1$. 

In the presence of $U$, we have diagonalized the Hamiltonian in the space of
many-body states in the DQD. We have found that the groundstate is either
spin singlet or doublet, depending on $\varphi_{1}$ and $\varphi_{2}$.
The supercurrent flows even in the doublet phase in our device.
The transition between the singlet and doublet phases enhances the SDE.
Its efficiency is almost the same as in the case of $U=0$.

For comparison, we examine a single QD connected to three superconducting
leads, depicted in Fig.\ \ref{Fig1:Model}(b), in Appendix C.
In the absence of $U$, (i) the groundstate energy takes the minimum at
$\varphi_1=\Phi$ in Eqs.\ (\ref{eq:phi_shift1a}) and (\ref{eq:phi_shift1b})
for a given $\varphi_{2}$. Unless $\varphi_{2} = 0$ or $\pi$,
$\Phi \ne 0$, $\pi$ and hence
$I_1$ flows at $\varphi_1=0$ (anomalous Josephson effect).
(ii) When the energy level $\varepsilon_0$ in the QD is tuned to
the Fermi level in the leads, the energies of the Andreev bound states
have zero points if $\Gamma_0$, $\Gamma_1$, and $\Gamma_2$ satisfy
the condition in Eq.\ (\ref{eq:triangle_inequality0}). Then,
the Dirac cones appear in the $\varphi_1-\varphi_2$ plane.
However, (iii) the device of single QD does not show the SDE. 
At least two QDs or two levels in a single QD are required for the SDE
as long as simple QDs are considered without the spin-orbit interaction,
etc.\cite{Cheng2023,Debnath2024}
Thus, our device in Fig.\ \ref{Fig1:Model}(a) should be one of the
minimal models to show the SDE.
(iv) In the presence of $U$, we set $\varepsilon_0=-U/2$.
Then, the groundstate is spin singlet or doublet.
The supercurrent $I_1$ is equivalent to that with $U=0$ 
in the singlet phase, whereas $I_1=0$ in the doublet phase.

In this paper, we have examined the effective Hamiltonian in
Eq.\ (\ref{eq:hamilotonian_DQD}), which is obtained by
the Schrieffer-Wolff transformation
to the second-order of the tunnel Hamiltonian $H_{\mathrm{T}}$.
Thus, our study is only applicable to the low-energy regime compared
with the superconducting gap; $|\varepsilon_j|$,
$U$, $\Gamma_{Rj}$, $\Gamma_{L;jj} \ll \Delta_0$ for $j=1,2$.
Otherwise, the higher-order terms with respect to $H_{\mathrm{T}}$ have to
be taken into account as in, e.g., Refs.\ \citen{Meng2009} and
\citen{Zalom2024} for a single QD.

Finally, we stress the merits of the QD device. The tunability of
the energy levels in the QDs enables the investigation of the
spectrum of the Andreev bound states in detail, artificial
creation of the Dirac cones in the $\varphi_1-\varphi_2$ plane,
and the realization of a large SDE.
This should lead to the deeper understanding of the physics of
Andreev bound states and molecules.

\begin{acknowledgments}
We appreciate fruitful discussions with Dr.\ Rui Sakano.
This work was partially supported by JSPS KAKENHI Grant Number JP20K03807.
\end{acknowledgments}

\appendix
\section{Bogoliubov Transformation and Dirac Points in the Absence of $U$}

\subsection{Bogoliubov transformation}

In the absence of $U$, the effective Hamiltonian is given by
Eq.\ (\ref{eq:Hamiltonian_Nambu}), or
\begin{equation}
  H_{\mathrm{eff}}^{\mathrm{DQD}}
  =\begin{pmatrix}
    d_{1\uparrow}^{\dag}, d_{1\downarrow}, d_{2\uparrow}^{\dag}, d_{2\downarrow}
    \end{pmatrix}
    \mathcal{H}
   \begin{pmatrix}
    d_{1\uparrow} \\
    d_{1\downarrow}^{\dag} \\
    d_{2\uparrow} \\
    d_{2\downarrow}^{\dag}
   \end{pmatrix} +\sum_{j=1, 2}\varepsilon_j
\label{eq:Hamiltonian_Nambu_r}
\end{equation}
with
\begin{equation}
  \mathcal{H}=
  \begin{pmatrix}
   \varepsilon_1 & \Gamma_{\mathrm{LAR}, 1} & 0 & \Gamma_{\mathrm{CAR}} \\
   \Gamma_{\mathrm{LAR},1}^* & -\varepsilon_1 & \Gamma_{\mathrm{CAR}}^* & 0 \\
   0 & \Gamma_{\mathrm{CAR}} & \varepsilon_2 & \Gamma_{\mathrm{LAR},2} \\
   \Gamma_{\mathrm{CAR}}^* & 0 & \Gamma_{\mathrm{LAR},2}^* & -\varepsilon_2
  \end{pmatrix}.
\end{equation}
The solution of the BdG equation, $\mathcal{H} {\bm u} = E {\bm u}$, is given by
$E=\pm E_1$, $\pm E_2$ ($0 \le E_1 \le E2$), where
\begin{equation}
  E_{1,2}=\sqrt{\frac{a \mp \sqrt{a^2-4b}}{2}},
\end{equation}
with
\begin{align}
    a &= \varepsilon_1^2+\varepsilon_2^2+|\Gamma_{\mathrm{LAR},1}|^2
         +|\Gamma_{\mathrm{LAR},2}|^2+2|\Gamma_{\mathrm{CAR}}|^2,
\\
    b &= \varepsilon_1^2\varepsilon_2^2
         +|\Gamma_{\mathrm{LAR},2}|^2\varepsilon_1^2
         +|\Gamma_{\mathrm{LAR},1}|^2\varepsilon_2^2
         +2\varepsilon_1\varepsilon_2|\Gamma_{\mathrm{CAR}}|^2
\nonumber \\
      &  +|\Gamma_{\mathrm{CAR}}|^4
          +|\Gamma_{\mathrm{LAR},1}|^2|\Gamma_{\mathrm{LAR},2}|^2
          - \left( \Gamma_{\mathrm{CAR}}^2\Gamma_{\mathrm{LAR},1}^*
            \Gamma_{\mathrm{LAR, 2}}^*+\mathrm{c.c.} \right).
\label{eq:BdGeq_b}
\end{align}
When ${\bm u}_n=(u_1,v_1,u_2,v_2)^T$ is the eigenfunction corresponding to
$E_n$,
$\tilde{\bm u}_n=(-v_1^*,u_1^*,-v_2^*,u_2^*)^T$ is that to $-E_n$.
The unitary matrix,
$U=({\bm u}_1, \tilde{\bm u}_1, {\bm u}_2, \tilde{\bm u}_2)$, satisfies
$P^{\dag}\tilde{H}P=\mathrm{diag}(E_1,-E_1,E_2,-E_2)$.
Thus the Bogoliubov transformation
\begin{equation}
    \label{eq:Bogoliubov}
    \begin{pmatrix}
        d_{1\uparrow} \\
         d_{1\downarrow}^{\dag} \\
          d_{2\uparrow} \\
           d_{2\downarrow}^{\dag}
    \end{pmatrix}
    =P
    \begin{pmatrix}
        \gamma_{1\uparrow} \\
         \gamma_{1\downarrow}^{\dag} \\
          \gamma_{2\uparrow} \\
           \gamma_{2\downarrow}^{\dag},
    \end{pmatrix}
\end{equation}
changes the Hamiltonian in Eq.\ (\ref{eq:Hamiltonian_Nambu_r}) to
\begin{equation}
  H_{\mathrm{eff}}^{\mathrm{DQD}}
    = \sum_{\sigma=\uparrow, \downarrow} \left(
      E_1 \gamma_{1\sigma}^{\dag}\gamma_{1\sigma}
     +E_2 \gamma_{2\sigma}^{\dag}\gamma_{2\sigma} \right)
     -\sum_{n=1,2}E_n+\sum_{j=1, 2}\varepsilon_j.
\label{eq:Hamiltonian_Bogoliubov}
\end{equation}
As a result, the groundstate energy is given by Eq.\ (\ref{eq:GSenergy0}).

Alternatively, $E_1$ and $E_2$ can be calculated using the retarded Green's
function in the Nambu form of $4 \times 4$ matrix.\cite{Pan2006}
Its trace possesses the poles at $E=\pm E_1 -i\delta$ and
$\pm E_2 -i\delta$ with an infinitesimally small $\delta$.

\begin{figure}
\centering
\includegraphics*[width=0.5\textwidth]{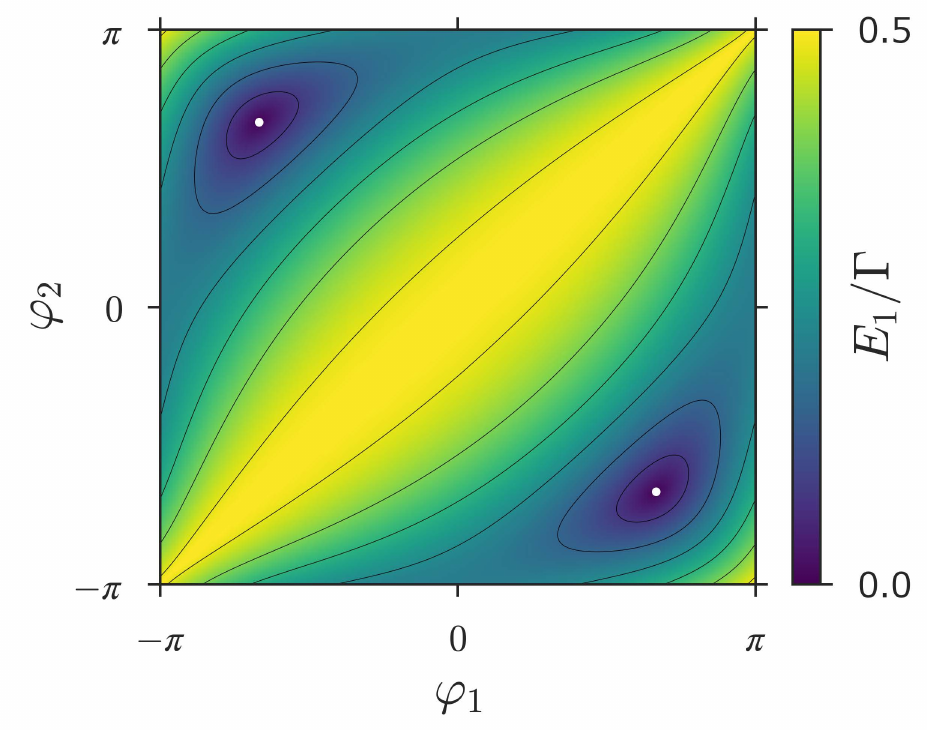}
\caption{(Color online)
Contour lines of the energy $E_1$ of Andreev bound state
in the $\varphi_1-\varphi_2$ plane, in the model in Fig.\ \ref{Fig1:Model}(a)
with $U=0$.
$\varepsilon_1=\varepsilon_2=0$,
$\Gamma_{L;11}=\Gamma_{R1}=\Gamma_{L;22}=\Gamma_{R2} \equiv \Gamma/2$,
and $p=1$. $E_1$ forms anisotropic Dirac cones around the zero points,
$\pm (2\pi/3, -2\pi/3)$, indicated by white dots.

}
\label{Fig9:DiracCone}
\end{figure}

\subsection{Dirac points}

Consider zero points of the spectrum of Andreev bound states.
The condition for $E_1=0$ is given by $b=0$ in Eq.\ (\ref{eq:BdGeq_b}).

First, we examine the case of 
$\Gamma_{L;11}=\Gamma_{R1}=\Gamma_{L;22}=\Gamma_{R2} \equiv \Gamma/2$.
Then,
\begin{align}
b &=
  \varepsilon_1^2\varepsilon_2^2
  +\frac{1}{2}\Gamma^2(1+\cos\varphi_2)\varepsilon_1^2
  +\frac{1}{2}\Gamma^2(1+\cos\varphi_1)\varepsilon_2^2
  +\frac{1}{2}\Gamma^2p^2\varepsilon_1\varepsilon_2
\nonumber \\
  & +\frac{\Gamma^4}{16}
   \left[ 4+p^4-2p^2+(4-2p^2)(\cos\varphi_1+\cos\varphi_2)
  +4\cos\varphi_1\cos\varphi_2-2p^2\cos(\varphi_1+\varphi_2) \right].
\label{eq:BdGeq_b2}
\end{align}
Here, we discuss a situation of  $|\varepsilon_1|=|\varepsilon_2|$.
In this case, the solutions of $b=0$
exist when $\varepsilon_1=-\varepsilon_2 \equiv \varepsilon_0$ and 
$|\varepsilon_0|<\Gamma p/2$.
There are two solutions, $(\varphi_1,\varphi_2)=
(\varphi_1^{(0)},\varphi_2^{(0)})$, where
\begin{align}
  \cos \varphi_1^{(0)} &=-1+\frac{1}{2}p^2-\frac{2\varepsilon_0^2}{\Gamma^2},
\label{eq:dirac1}
\\
\varphi_2^{(0)} &=-\varphi_1^{(0)}.
\label{eq:dirac2}
\end{align}
Around these zero-points, $E_1(\varphi_1,\varphi_2)$ behaves as
\begin{align}
   E_1(\varphi_1^{(0)} & +\delta\varphi_1, \varphi_2^{(0)} +\delta\varphi_2)
    \nonumber \\
    &= \frac{\Gamma}{4}
    \sqrt{1+\left[p^2\left(\frac{4 \varepsilon_0^2}{\Gamma^2p^2}\right)^2
    +(4-2p^2) \left(\frac{4 \varepsilon_0^2}{\Gamma^2p^2}\right)+p^2-2 \right]
    \cos\theta\sin\theta}
    \delta\varphi,
\label{eq:diracE1}
\end{align}
where 
$\delta\varphi_1=\delta \varphi \cos\theta$ and
$\delta\varphi_2=\delta\varphi \sin\theta$ ($\delta\varphi > 0$).
This indicates an anisotropic Dirac cone.

When $p=1$ and $\varepsilon_0=0$,
$E_1(\varphi_1,\varphi_2)$ is shown
in Fig.\ \ref{Fig9:DiracCone} by the contour lines.
Equations (\ref{eq:dirac1})--(\ref{eq:diracE1})
yield $\varphi_1^{(0)}=-\varphi_2^{(0)}=\pm 2\pi/3$ and
$E_1 \simeq (\Gamma/4) \sqrt{1-\cos\theta\sin\theta}\delta \varphi$
around the Dirac points.
When $\delta\varphi_2=0$ ($\theta=0,\pi$),
$E_1(\varphi_1^{(0)} \pm \delta\varphi, \varphi_2^{(0)})
\simeq (\Gamma/4) \delta\varphi$,
and thus $\partial E_1/ \partial \varphi_1=\pm \Gamma/4$
at $\varphi_1=\varphi_1^{(0)} \pm 0$ and
$\varphi_2=\varphi_2^{(0)}$.

Second, we derive the condition for the existence of Dirac points when
$\varepsilon_1=\varepsilon_2=0$ with arbitrary $\Gamma_{L;11}$,
$\Gamma_{R1}$, $\Gamma_{L;22}$, $\Gamma_{R2}$, and $p$.
Then,
\begin{align}
  b &=\left| \Gamma_{\mathrm{CAR}}^2
    -\Gamma_{\mathrm{LAR}, 1}\Gamma_{\mathrm{LAR}, 2} \right|^2
    \nonumber \\
    &=\left| \Gamma_{L;11}\Gamma_{L;22}(1-p^2)
    +\Gamma_{R1}\Gamma_{L;22}e^{i\varphi_1}
    +\Gamma_{L;11}\Gamma_{R2}e^{i\varphi_2}
    +\Gamma_{R1}\Gamma_{R2}e^{i(\varphi_1+\varphi_2)} \right|^2.
\label{eq:equation_of_b}
\end{align}
The Dirac points exist when $b=0$ at some $(\varphi_1,\varphi_2)$.
For $p=1$, the solutions exist when 
$\Gamma_{R1}\Gamma_{L;22}$, $\Gamma_{L;11}\Gamma_{R2}$, and
$\Gamma_{R1}\Gamma_{R2}$ can make a triangle, or
\begin{equation}
|\Gamma_{R1}\Gamma_{L;22}-\Gamma_{L;11}\Gamma_{R2}| <
\Gamma_{R1}\Gamma_{R2} < \Gamma_{R1}\Gamma_{L;22}+\Gamma_{L;11}\Gamma_{R2}
\label{eq:triangle_inequality}
\end{equation}
is satisfied. This condition is usually realized if the $\Gamma$'s are
in the same order of magnitude.

\section{Hamiltonian in Space of Many-Body States}

In the presence of $U$, we diagonalize the effective Hamiltonian
$H_{\mathrm{eff}}^{\mathrm{DQD}}$ in Eq.\ (\ref{eq:hamilotonian_DQD})
in the space of many-body states in the DQD.
The Hamiltonian is block-diagonalized
into three subspaces with the total spin of $S=0$, $1/2$, and $1$.

For spin singlet ($S=0$), there are five states,
$\ket{0}$, $\ket{s}$ in Eq.\ (\ref{eq:singlet_s}),
$\ket{1\uparrow, 1\downarrow}$,
$\ket{2\uparrow, 2\downarrow}$, and
$\ket{1\uparrow, 1\downarrow, 2\uparrow, 2\downarrow}$.
Using these states as a basis set, the Hamiltonian is represented by
\begin{equation}
\label{eq:matrix_singlet}
   \begin{pmatrix}
       0 & \sqrt{2}\Gamma_{\mathrm{CAR}}^* & \Gamma_{\mathrm{LAR}, 1}^* &
       \Gamma_{\mathrm{LAR}, 2}^* & 0 \\ 
       \sqrt{2}\Gamma_{\mathrm{CAR}} & \varepsilon_1+\varepsilon_2 & 0 & 0 &
       -\sqrt{2}\Gamma_{\mathrm{CAR}}^* \\
       \Gamma_{\mathrm{LAR},1} & 0 & 2\varepsilon_1+U & 0 &
       \Gamma_{\mathrm{LAR}, 2}^* & \\
       \Gamma_{\mathrm{LAR},2} & 0 & 0 & 2\varepsilon_2+U &
       \Gamma_{\mathrm{LAR}, 1}^* \\
       0 & -\sqrt{2}\Gamma_{\mathrm{CAR}} & \Gamma_{\mathrm{LAR}, 2} &
      \Gamma_{\mathrm{LAR}, 1} & 2(\varepsilon_1+\varepsilon_2+U)
   \end{pmatrix}.
\end{equation}
When $U=0$, the eigenenergies of Eq.\ (\ref{eq:matrix_singlet}) are
$\varepsilon_1+\varepsilon_2\pm E_1 \pm E_2$ and $\varepsilon_1+\varepsilon_2$,
where $E_1$ and $E_2$ are the eigenenergies of the BdG equation in Sect.\ A.1.
The lowest energy, $\varepsilon_1+\varepsilon_2-E_1-E_2$, coincides with
$E_{\mathrm{GS}}$ in Eq.\ (\ref{eq:GSenergy0}). If the groundstate is denoted
by $\ket{\mathrm{GS}}$, the other states are written as
$\gamma_{1\uparrow}^{\dag}\gamma_{1\downarrow}^{\dag} \ket{\mathrm{GS}}$,
$\gamma_{2\uparrow}^{\dag}\gamma_{2\downarrow}^{\dag} \ket{\mathrm{GS}}$,
$\gamma_{1\uparrow}^{\dag}\gamma_{1\downarrow}^{\dag}
 \gamma_{2\uparrow}^{\dag}\gamma_{2\downarrow}^{\dag} \ket{\mathrm{GS}}$,
and $(\gamma_{1\uparrow}^{\dag}\gamma_{2\downarrow}^{\dag}-
     \gamma_{1\downarrow}^{\dag}\gamma_{2\uparrow}^{\dag})
    \ket{\mathrm{GS}}/\sqrt{2}$.

For spin doublet ($S=1/2$), there are four states,
$\ket{1\uparrow}$, $\ket{2\uparrow}$,
$\ket{1\uparrow, 2\uparrow, 2\downarrow}$,
$\ket{1\uparrow, 1\downarrow, 2\uparrow}$, with $S_z=1/2$,
and four states,
$\ket{1\downarrow}$, $\ket{2\downarrow}$,
$\ket{1\downarrow, 2\uparrow, 2\downarrow}$,
$\ket{1\uparrow, 1\downarrow, 2\downarrow}$, with $S_z=-1/2$. 
The representation matrix of the Hamiltonian is given by
\begin{equation}
\label{eq:matrix_doublet}
  \begin{pmatrix}
   \varepsilon_1 & 0 & \Gamma_{\mathrm{LAR}, 2}^* & -\Gamma_{\mathrm{CAR}}^*
  \\
   0 & \varepsilon_2 & -\Gamma_{\mathrm{CAR}}^* & \Gamma_{\mathrm{LAR}, 1}^*
  \\
   \Gamma_{\mathrm{LAR}, 2} & -\Gamma_{\mathrm{CAR}} &
   \varepsilon_1+2\varepsilon_2+U & 0 \\
   -\Gamma_{\mathrm{CAR}} & \Gamma_{\mathrm{LAR}, 1} & 0 &
   2\varepsilon_1+\varepsilon_2+U
  \end{pmatrix}
\end{equation}
for both $S_z=1/2$ and $-1/2$.
When $U=0$, the eigenenergies are $\varepsilon_1+\varepsilon_2\pm E_1$
and $\varepsilon_1+\varepsilon_2\pm E_2$. The eigenstates are 
$\gamma_{1\sigma}^{\dag}\ket{\mathrm{GS}}$,
$\gamma_{2\sigma}^{\dag}\ket{\mathrm{GS}}$,
$\gamma_{1\sigma}^{\dag}\gamma_{2\uparrow}^{\dag}\gamma_{2\downarrow}^{\dag}
\ket{\mathrm{GS}}$,
and
$\gamma_{1\uparrow}^{\dag}\gamma_{1\downarrow}^{\dag}\gamma_{2\sigma}^{\dag}
\ket{\mathrm{GS}}$
with $\sigma=\uparrow$ or $\downarrow$.

For spin triplet ($S=1$), there are three states,
$\ket{1\uparrow, 2\uparrow}$, $\ket{t}$ in Eq.\ (\ref{eq:triplet_t}),
and $\ket{1\downarrow, 2\downarrow}$.
They are eigenstates of the Hamiltonian with eigenenergy
$\varepsilon_1+\varepsilon_2$ even in the presence of $U$.
They are given by
$\gamma_{1\uparrow}^{\dag}\gamma_{2\uparrow}^{\dag} \ket{\mathrm{GS}}$,
$\gamma_{1\downarrow}^{\dag}\gamma_{2\downarrow}^{\dag} \ket{\mathrm{GS}}$,
and $(\gamma_{1\uparrow}^{\dag}\gamma_{2\downarrow}^{\dag}+
      \gamma_{1\downarrow}^{\dag}\gamma_{2\uparrow}^{\dag})
    \ket{\mathrm{GS}}/\sqrt{2}$,
where $\ket{\mathrm{GS}}$ is the spin-singlet groundstate in the absence of
$U$.

\section{Single QD with Three Superconducting Leads}

Here, we examine a single QD connected to three superconducting leads, as
depicted in Fig.\ \ref{Fig1:Model}(b). By the same procedure as in
Sect.\ 2.1, we derive the effective Hamiltonian,
\begin{equation}
  \label{eq:hamilotonian_QD}
  H_{\mathrm{eff}}^{\mathrm{QD}}=H_{\mathrm{dot}}+H_{\mathrm{AR}}
\end{equation}
with
\begin{align}
  H_{\mathrm{dot}} &=\sum_{\sigma}\varepsilon_0 d_{\sigma}^{\dag}d_{ \sigma}
  + Un_{\uparrow}n_{\downarrow},
\label{eq:H_SQD}
\\
  H_{\mathrm{AR}} & =
  \Gamma_{\mathrm{AR}} d_{\uparrow}^{\dag}d_{\downarrow}^{\dag}+
  \mathrm{h.c.},
\label{eq:H_AR}
\end{align}
where
\begin{equation}
\Gamma_{\mathrm{AR}}=-\Gamma_0e^{i\phi_0}-\Gamma_1e^{i\phi_1}-
\Gamma_2e^{i\phi_2}.
\label{eq:Gamma_AR}
\end{equation}
Here, $\varepsilon_0$ and $U$ are the energy level and electron-electron
interaction in the QD, respectively.
$n_{\sigma}=d_{\sigma}^{\dag} d_{\sigma}$ is the number operator with spin
$\sigma$ in the QD.
$\Gamma_j$ is the level broadening due to the tunnel
coupling to lead $j$, which is defined by a similar equation to
Eq.\ (\ref{eq:linewidth_fn_R}). $\phi_j$ is the superconducting phase
in lead $j$. We introduce the phase differences
$\varphi_1=\phi_0-\phi_1$ and $\varphi_2=\phi_0-\phi_2$.

\subsection{In case of $U=0$}

We begin with the case of $U=0$.
The energies of the Andreev bound states are $\pm E_1$, where
\begin{equation}
E_1=\sqrt{\varepsilon_0^2+|\Gamma_{\mathrm{AR}}|^2}.
\label{eq:QDE1}
\end{equation}
The groundstate energy is given by $E_{\mathrm{GS}}=-E_1+\varepsilon_0$.
The supercurrent to lead $j$ ($=1,2$) is obtained by
\begin{equation}
I_j(\varphi_1,\varphi_2)=\frac{2e}{\hbar}
\pdv{E_{\mathrm{GS}}}{\varphi_j}
=-\frac{2e}{\hbar} \pdv{E_1}{\varphi_j}
\label{eq:supercurrentQD1}
\end{equation}
at $T=0$.

\begin{figure}
\centering
\includegraphics*[width=0.6\textwidth]{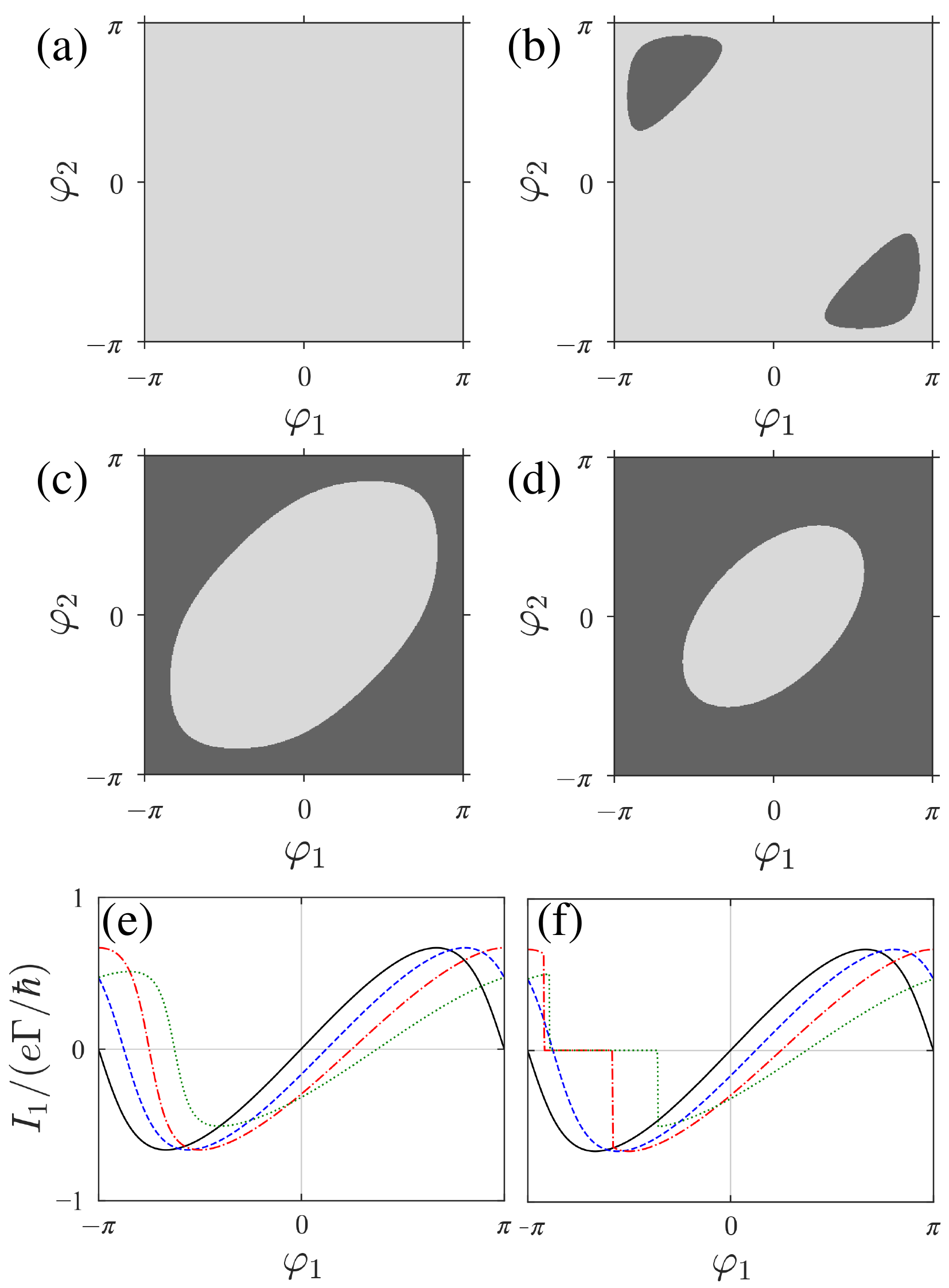}
\caption{(Color online)
(a)--(d) Spin of the groundstate and (e), (f)
supercurrent $I_1$ to lead $1$ in the model in Fig.\ \ref{Fig1:Model}(b).
$\varepsilon_0=-U/2$ and
$\Gamma_{0}=\Gamma_{1}=\Gamma_{2} \equiv \Gamma/3$.
The spin singlet (doublet) regions are indicated by light gray
(dark gray) in the $\varphi_1-\varphi_2$ plane for (a) $U/\Gamma=0$,
(b) $0.5$, (c) $1$, and (d) $1.5$.
The supercurrent $I_1$ is plotted as a function of $\varphi_1$ when
$\varphi_2$ is fixed at $0$ (solid line), $\pi/4$ (broken line),
$\pi/2$ (dash-dotted line), and $3\pi/4$ (dotted line),
for (e) $U/\Gamma=0$ and (f) $0.5$. In panel (f), $I_1=0$ in the
doublet region.
}
\label{Fig10:singleQD}
\end{figure}

Let us examine $E_1$ in Eq.\ (\ref{eq:QDE1}). 
$|\Gamma_{\mathrm{AR}}|^2$ is rewritten as
\begin{align}
    |\Gamma_{\mathrm{AR}}|^2=
    & \Gamma_0^2+ \Gamma_1^2+ \Gamma_2^2+2\Gamma_0\Gamma_2\cos\varphi_2
\nonumber \\
    & +2\Gamma_1\sqrt{\Gamma_0^2+\Gamma_2^2+2\Gamma_0\Gamma_2\cos\varphi_2}
     \cos(\varphi_1-\Phi),
\end{align}
where $\Phi$ is given by
\begin{align}
  \cos\Phi=\frac{\Gamma_0+\Gamma_2\cos\varphi_2}
  {\sqrt{\Gamma_0^2+\Gamma_2^2+2\Gamma_0\Gamma_2\cos\varphi_2}},
\label{eq:phi_shift1a}
\\
  \sin\Phi=\frac{\Gamma_2\sin\varphi_2}
 {\sqrt{\Gamma_0^2+\Gamma_2^2+2\Gamma_0\Gamma_2\cos\varphi_2}}.
\label{eq:phi_shift1b}
\end{align}
For a given $\varphi_2$, the groundstate energy $E_{\mathrm{GS}}$
is the minimum at $\varphi_1=\Phi$.
As a result, the supercurrent $I_1$ flows at $\varphi_1=0$ unless
$\Phi=0,\pi$ or $\varphi_2=0,\pi$ (anomalous Josephson effect),
as seen in Fig.\ \ref{Fig10:singleQD}(e).
Note that $\Phi$ can be controlled by $\varphi_2$ in this device.

Since $E_{\mathrm{GS}}$ is an even function of $(\varphi_1-\Phi)$,
the maximum and minimum of $I_1$ are identical in magnitude,
$I_{\mathrm{c}}^-=I_{\mathrm{c}}^+$. Hence no SDE is observed.

We investigate the zero points of the energies of the Andreev
bound states, $\pm E_1$ in Eq.\ (\ref{eq:QDE1}).
If $\varepsilon_0=0$, the solutions of $|\Gamma_{\mathrm{AR}}|=0$
exist when the triangle inequality
\begin{equation}
|\Gamma_1-\Gamma_2| < \Gamma_0 < \Gamma_1+\Gamma_2
\label{eq:triangle_inequality0}
\end{equation}
is fulfilled.\cite{Zalom2024,com2}
Then, there are two solutions given by
\begin{align}
  \cos\varphi_1 &
  =\frac{\Gamma_2^2-\Gamma_1^2-\Gamma_0^2}{2\Gamma_0\Gamma_1},
\\
   \cos\varphi_2 &
  =\frac{\Gamma_1^2-\Gamma_2^2-\Gamma_0^2}{2\Gamma_0\Gamma_2},
\\
  \sin\varphi_2 &
  =-\frac{\Gamma_1}{\Gamma_2}\sin\varphi_1.
\end{align}
When $\Gamma_0=\Gamma_1=\Gamma_2$, they are
$(\varphi_1,\varphi_2)=(2\pi/3,-2\pi/3)$ and $(-2\pi/3,2\pi/3)$.

When $\varepsilon_0=0$, 
we expand $E_1=|\Gamma_{\mathrm{AR}}|$ around a zero point,
$(\varphi_1^{(0)}, \varphi_2^{(0)})$.
For $\varphi_1=\varphi_1^{(0)} +\delta\varphi_1$ and
$\varphi_2=\varphi_2^{(0)} +\delta\varphi_2$,
\begin{equation}
  E_1 \simeq \sqrt{\Gamma_1^2\cos^2\theta+\Gamma_2^2\sin^2\theta
    -\frac{1}{2}(\Gamma_1^2+\Gamma_2^2-\Gamma_0^2)\sin 2\theta}
    \delta\varphi,
\end{equation}
where $\delta\varphi_1=\delta \varphi \cos\theta$ and
$\delta\varphi_2=\delta\varphi \sin\theta$ ($\delta \varphi > 0$).
This indicates an anisotropic Dirac cone.
When $\delta\varphi_2=0$ ($\theta=0,\pi$),
$E_1(\varphi_1^{(0)} \pm \delta\varphi, \varphi_2^{(0)})
\simeq \Gamma_1 \delta\varphi$,
and thus $\partial E_1/ \partial \varphi_1=\pm \Gamma_1$
at $\varphi_1=\varphi_1^{(0)} \pm 0$ and $\varphi_2=\varphi_2^{(0)}$.

\subsection{In case of $U \ne 0$}

In the presence of $U$, the groundstate is spin singlet or doublet in the
device of single QD. We rewrite $H_{\mathrm{dot}}$ in
Eq.\ (\ref{eq:H_SQD}) as
\begin{equation}
  H_{\mathrm{dot}}=\sum_{\sigma}\xi_0 d_{\sigma}^{\dag}d_{\sigma}
  +\frac{U}{2}\left(\sum_{\sigma}n_{\sigma}-1\right)^2-\frac{U}{2},
\label{eq:H_SQD2}
\end{equation}
where $\xi_0=\varepsilon_0+(U/2)$. For the spin-singlet states,
which are given by linear combinations of
$\ket{0}$ and $\ket{\uparrow, \downarrow}$, the second and
third terms are cancelled with each other on the right
side of Eq.\ (\ref{eq:H_SQD2}).
The lowest energy is
\begin{equation}
E_{\mathrm{singlet}}(\varphi_1,\varphi_2)=
- \sqrt{\xi_0^2+|\Gamma_{\mathrm{AR}}|^2} + \xi_0
\end{equation}
with $\Gamma_{\mathrm{AR}}$ in Eq.\ (\ref{eq:Gamma_AR}).
This equation is just obtained by replacing $\varepsilon_0$ with
$\xi_0$ in the groundstate energy for $U=0$.\cite{Meden2019}

The spin-doublet states ($\ket{\uparrow}$, $\ket{\downarrow}$)
have the energy of $E_{\mathrm{doublet}}=\varepsilon_0=\xi_0-(U/2)$
since $H_{\mathrm{AR}}$ is irrelevant in $H_{\mathrm{eff}}^{\mathrm{QD}}$
in Eq.\ (\ref{eq:hamilotonian_QD}).
In the doublet phase, no supercurrent flows because 
$E_{\mathrm{doublet}}$ does not depend on $\varphi_1$.

The groundstate energy $E_{\mathrm{GS}}$ is determined by the minimum
between $E_{\mathrm{singlet}}$ and $E_{\mathrm{doublet}}$.
In Figs.\ \ref{Fig10:singleQD}(a)--(d), we show the phase diagram
of the single QD in the
$\varphi_1-\varphi_2$ plane for several values of $U$.
We set $\xi_0=0$.
The singlet phase always appears for $U=0$ while both singlet
and doublet phases are seen for $U \ne 0$.

Fixing at $\xi_0=0$,
we evaluate the supercurrent to lead $1$, $I_1$, using the first equation
in Eq.\  (\ref{eq:supercurrentQD1}).
In Figs.\ \ref{Fig10:singleQD}(e) and (f),
the supercurrent $I_1$ is depicted as a function of $\varphi_1$
for a given $\varphi_2$ when (e) $U=0$ and (f) $U/\Gamma=0.5$ with
$\Gamma=\Gamma_0+\Gamma_1+\Gamma_2$.
In the presence of $U$,
the current is equivalent to that for $U=0$ in the singlet phase,
whereas $I_1=0$ in the doublet phase.


\begin{thebibliography}{99}
\bibitem{Ando2020}
F.\ Ando, Y.\ Miyasaka, T.\ Li, J.\ Ishizuka, T.\ Arakawa, Y.\ Shiota,
T.\ Moriyama, Y.\ Yanase, and T. Ono, Nature\ \textbf{584}, 373 (2020).
%
\bibitem{Hu2007}
L.\ Hu, C.\ Wu, and X.\ Dai,
Phys.\ Rev.\ Lett.\ \textbf{99}, 067004 (2007).
\bibitem{Reynoso2008}
A.\ A.\ Reynoso, G.\ Usaj, C.\ A.\ Balseiro, D.\ Feinberg, and M.\ Avignon,
Phys.\ Rev.\ Lett.\ \textbf{101}, 107001 (2008).
\bibitem{Zazunov2009}
A.\  Zazunov, R.\ Egger, T.\ Jonckheere, and T.\  Martin,
Phys.\ Rev.\ Lett.\ \textbf{103}, 147004 (2009).
%
\bibitem{Yokoyama2013}
T.\ Yokoyama, M.\ Eto, and Yu.\ V.\ Nazarov,
J.\ Phys.\ Soc.\ Jpn.\ \textbf{82}, 054703 (2013).
\bibitem{Yokoyama2014}
T.\ Yokoyama, M.\ Eto, and Yu.\ V.\ Nazarov,
Phys.\ Rev.\ B \textbf{89}, 195407 (2014).
%
 \bibitem{Baumgartner2022}
    C.\  Baumgartner, L.\ Fuchs, A.\  Costa, S.\  Reinhardt, S.\ Gronin, G.\ C.\ Gardner,
   T.\ Lindemann, M.\ J.\ Manfra, P.\ E.\ Faria,\ Jr., D.\ Kochan, J.\ Fabian,
    N.\ Paradiso, and C.\ Strunk, Nat.\ Nanotechnol.\ \textbf{17}, 39 (2022).
\bibitem{Wu2022}
    H.\ Wu, Y.\ Wang, Y.\ Xu, P.\ K.\ Sivakumar, C.\ Pasco, U.\ Filippozzi, S.\ S.\ P.\ Parkin, Y.-J.\ Zeng, 
    T.\ McQueen, and M.\ N.\ Ali, Nature\ \textbf{604}, 653 (2022).
\bibitem{Pal2022}
B.\ Pal, A.\ Chakraborty, P.\ K.\ Sivakumar, M.\ Davydova, A.\ K.\ Gopi, A.\ K.\ Pandeya,
J.\ A.\ Krieger, Y.\ Zhang, M.\ Date, S.\ Ju, N.\ Yuan, N.\ B.\ M.\ Schr\"oter, L.\ Fu, and S.\ S.\ P.\ Parkin,
Nat.\ Phys.\ \textbf{18}, 1228 (2022).
 \bibitem{Costa2023}
A.\ Costa, C.\ Baumgartner, S.\ Reinhardt, J.\ Berger, S.\ Gronin, G.\ C.\ Gardner, T.\ Lindemann,
M.\ J.\ Manfra, J.\ Fabian, D.\ Kochan, N.\ Paradiso, and C.\ Strunk,
Nat.\ Nanotechnol.\ \textbf{18}, 1266 (2023).
\bibitem{Banerjee2023}
 A.\ Banajee, M.\ Geier, M.\ A.\ Rahman, C.\ Thomas, T.\ Wang, M.\ J.\ Manfra, K.\ Flensberg,
 and C.\ M.\ Marcus, Phys.\ Rev.\ Lett.\ \textbf{131}, 196301 (2023).
%
\bibitem{Ciaccia2023}
  C.\ Ciaccia, R.\ Haller, A.\ C.\ C.\ Drachmann, T.\ Lindemann, M.\ J.\ Manfra,
  C.\ Schrade, and C.\ Schr\"{o}nenberger, Phys.\ Rev.\ Res.\ \textbf{5}, 033131 (2023).
\bibitem{Greco2023}
  A.\ Greco, Q.\ Pichard, and F.\ Giazotto, Appl.\ Phys.\ Lett.\ \textbf{123}, 092601 (2023).
%
\bibitem{Gupta2023}
  M.\ Gupta, G.\ V.\ Graziano, M.\ Pendharkar, J.\ T.\ Dong, C.\ P.\ Dempsey, C.\ Palmstrøm,
  and V.\ S.\ Pribiag, Nat.\ Commun.\ \textbf{14}, 3078 (2023).
\bibitem{Coraiola2023}
   M.\ Coraiola, D.\ Z.\ Haxwell, D.\ Sabonis, H.\ Weisbrich, A.\ E.\ Svetogorov,
   M.\ Hinderling, S.\ C. ten\ Kate, E.\ Cheah, F.\ Krizek, R.\ Schott, W.\ Wegscheider,
   J.\ C.\ Cuevas, W.\ Belzig, and F.\ Nichele, Nat.\ Commun.\ \textbf{14}, 6784 (2023).
%
\bibitem{Matsuo2023_1}
  S.\ Matsuo, T.\ Imoto, T.\ Yokoyama, Y.\ Sato, T.\ Lindemann, S.\ Gronin,
  G.\ C.\ Gardner, M.\ J.\ Manfra, and S.\ Tarucha, Nat.\ Phys.\ \textbf{19}, 1636 (2023).
\bibitem{Matsuo2023_2}
  S.\ Matsuo, T.\ Imoto, T.\ Yokoyama, Y.\ Sato, T.\ Lindemann, S.\ Gronin, G.\ C.\ Gardner,
  M.\ J.\ Manfra, and S.\ Tarucha, Sci.\ Adv.\ \textbf{9}, eadj3698 (2023).
%
\bibitem{Kornich2019}
  V.\ Kornich, H.\ S.\ Barakov, and Yu.\ V.\ Nazarov,
Phys.\ Rev.\ Res.\ \textbf{1}, 033004 (2019). 
\bibitem{Kornich2020}
  V.\ Kornich, H.\ S.\ Barakov, and Yu.\ V.\ Nazarov,
Phys.\ Rev.\ B \textbf{101}, 195430 (2020). 
\bibitem{Matsuo2022}
  S.\ Matsuo, J.\ S.\ Lee, C.-Y.\ Chamg. Y.\ Sato, K.\ Ueda, C.\ J.\ Palmstrøm, and S.\ Tarucha,
  Commun\ Phys.\ \textbf{5}, 221 (2022). 
\bibitem{Pillet2023}
  J.-D.\ Pillet, S.\ Annabi, A.\ Peugeot, H.\ Riechert, E.\ Arrighi, J.\ Griesmar, and L.\ Bretheau,
  Phys.\ Rev.\ Res.\ \textbf{5}, 033199 (2023).
\bibitem{Matsuo2023_3}
  S.\ Matsuo, T.\ Imoto, T.\ Yokoyama, Y.\ Sato, T.\ Lindemann, S.\ Gronin, G.\ C.\ Gardner, S.\ Nakosai,
  Y.\ Tanaka, M.\ J.\ Manfra, and S.\ Tarucha, Nat.\ Commun.\ \textbf{14}, 8271 (2023).
%
\bibitem{Meden2019}
V.\ Meden, J.\ Phys.: Condens.\ Matter  \textbf{31}, 163001 (2019).
\bibitem{Dam2006}
J.\ A.\ van Dam, Yu.\ V.\ Nazarov, E.\  P.\ A.\ M.\ Bakkers, S.\ De Franceschi,
 and L.\ P.\ Kouwenhoven, Nature\ \textbf{442}, 667 (2006).
%
\bibitem{Choi2000}
M.-S.\ Choi, C.\ Bruder, and D.\ Loss,
Phys.\ Rev.\ B \textbf{62}, 13569 (2000).
\bibitem{Pan2006}
  H.\ Pan and T.-H.\ Lin, Phys.\ Rev.\ B \textbf{74}, 235312 (2006).
\bibitem{Wang2011}
  Z.\ Wang and X.\ Hu, Phys.\ Rev.\ Lett.\ \textbf{106}, 037002 (2011).
\bibitem{Recher2001}
  P.\ Recher, E.\ V.\ Sukhorukov, and D.\ Loss, Phys.\ Rev.\ B \textbf{63}, 165314 (2001).
\bibitem{Deacon2015}
  R.\ S.\ Deacon, A.\ Oiwa, J.\ Sailer, S.\ Baba, Y.\ Kanai, K.\ Shibata, K.\ Hirakawa,
  and S.\ Tarucha, Nat.\ Commun.\ \textbf{6}, 7446 (2015).
%
\bibitem{Leijnse2012}
  M\ Leijnse and K.\ Flensberg, Phys.\ Rev.\ B \textbf{86}, 134528 (2012).
\bibitem{Dvir2023}
  T.\ Dvir, G.\ Wang, N.\ van\ Loo, C.-X.\ Liu, G.\ P.\ Mazur, A.\ Bordin,
  S.\ L.\ D.\ ten\ Haaf, J.-Y.\ Wang, D.\ van\ Driel, F.\ Zatelli, X.\ Li,
  F.\ K.\ Malinowski, S.\ Gazibegovic, G.\ Badawy, E.\ P.\ A.\ M.\ Bakkers,
  M.\ Wimmer, and L.\ P.\ Kouwenhoven, Nature\ \textbf{614}, 445 (2023).
\bibitem{Bordin2023}
  A.\ Bordin, G.\ Wang, C.-X.\ Liu, S.\ L.\ D.\ ten Haaf, N.\ van\ Loo, G.\ P.\ Mazur,
  D.\ Xu, D.\ van, Driel, F.\ Zatelli, S.\ Gazibegovic,
  G.\ Badawy, E.\ P.\ A.\ M.\ Bakkers, M.\ Wimmer, L.\ P.\ Kouwenhoven, and T.\ Dvir,
  Phys.\ Rev.\ X.\ \textbf{13}, 031031 (2023).
%
\bibitem{Yokoyama2015}
  T.\ Yokoyama and Yu.\ V.\ Nazarov, Phys.\ Rev.\ B \textbf{92}, 155437 (2015).
\bibitem{Riwar2016}
  R.-P.\ Riwar, M.\ Houzet, J.\ S.\ Meyer, and Yu.\ V.\ Nazarov,
  Nat.\ Commun.\ \textbf{7}, 11167 (2016).
\bibitem{Klees2020}
  R.\ L.\ Klees, G.\ Rastelli, J.\ C.\ Cuevas, and W.\ Belzig,
  Phys.\ Rev.\ Lett.\ \textbf{124}, 197002 (2020). 
%
\bibitem{Cheng2023}
Q.\ Cheng and Q.-F.\ Sun, Phys.\ Rev.\ B \textbf{107}, 184511 (2023).
\bibitem{Debnath2024}
D.\ Debnath and P.\ Dutta, Phys.\ Rev.\ B \textbf{109}, 174511 (2024).
%
\bibitem{Kubo2006}
T.\ Kubo, Y.\ Tokura, T.\ Hatano, and S.\ Tarucha,
Phys.\ Rev.\ B \textbf{74}, 205310 (2006). 
\bibitem{Zhang2022}
Y.\ Zhang, R.\ Sakano, and M.\ Eto,
J.\ Phys.\ Soc.\ Jpn.\ \textbf{91}, 014703 (2022).
\bibitem{Zhang2024}
Y.\ Zhang, M.\ Kato, R.\ Sakano, and M.\ Eto,
J.\ Phys.\ Soc.\ Jpn.\ \textbf{93}, 024702 (2024).
\bibitem{Eto2020}
M.\ Eto and R.\ Sakano, Phys.\ Rev.\ B \textbf{102}, 245402 (2020).
%
\bibitem{Bravyi2011}
S.\ Bravyi, D.\ P.\ DiVincenzo, and D.\ Loss,
Ann.\ Phys.\ \textbf{326}, 2793 (2011).
\bibitem{Scherubl2019}
  Z.\ Scher\"{u}bl, A.\ P\'{a}lyi, and S.\ Csonka,
  Beilstein J.\ Nanotechnol.\ \textbf{10}, 363 (2019).
\bibitem{Spethmann2024}
  M.\ Spethmann, S.\ Bosco, A.\ Hofmann, J.\ Klinovaja, and D.\ Loss,
  Phys.\ Rev.\ B \textbf{109}, 085303 (2024).
\bibitem{com1}
Our numerical study indicates that $\varphi_2$ at the Dirac points
almost coincides with $\varphi_2$ at the maximum of
$I_{\mathrm{c}}^{\pm}$ when $|\varepsilon_0|/\Gamma
\lesssim 0.4$.
\bibitem{Meng2009}
T.\ Meng and S.\ Florens, and P.\ Simon,
Phys.\ Rev.\ B \textbf{79}, 224521 (2009).
\bibitem{Zalom2024}
  P.\ Zalom, M.\ \u{Z}onda, and T.\ Novotn\'{y}, Phys.\ Rev.\ Lett.\
  \textbf{132}, 126505 (2024).
\bibitem{com2}
In the two-terminal geometry ($\Gamma_2=0$),
the energies of Andreev bound state have a zero point when
$\varepsilon_0=0$ and $\Gamma_1=\Gamma_2$.
Regarding the condition for $\Gamma$'s, the zero points
are easily realized in the three-terminal geometry.

\end{thebibliography}
\end{document}